\documentclass[nofootinbib,prb,twocolumn,amsmath,amssymb,aps,]{revtex4-1}
\usepackage[utf8]{inputenc}
\usepackage{graphicx}
\usepackage{adjustbox}
\usepackage{dcolumn}
\usepackage{bm}
\usepackage{hyperref}
\usepackage{mathtools}
\usepackage{physics}
\hypersetup{
     colorlinks   = true,
     linkcolor    = blue,
     citecolor    = blue
}
\usepackage[all]{hypcap}
\begin{document}

\title{Scrambling in the Dicke model}
\author{Yahya Alavirad$^1$}
\author{Ali Lavasani$^1$}

\affiliation{
$^1$Department of Physics, Condensed Matter theory center and the Joint Quantum Institute, University of Maryland, College Park, MD 20742, USA}

\date{\today}

\begin{abstract}
The scrambling rate $\lambda_L$ associated with the exponential growth of out-of-time-ordered correlators can be used to characterize quantum chaos. Here we use the Majorana Fermion representation of spin $1/2$ systems to study quantum chaos in the Dicke model. We take the system to be in thermal equilibrium and compute $\lambda_L$ throughout the phase diagram to leading order in $1/N$. We find that the chaotic behavior is strongest close to the critical point. At high temperatures $\lambda_L$ is nonzero over an extended region that includes both the normal and super-radiant phases. At low temperatures $\lambda_L$ is nonzero in (a) close vicinity of the critical point and (b) a region within the super-radiant phase. In the process we also derive a new effective theory for the super-radiant phase at finite temperatures. Our formalism does not rely on the assumption of total spin conservation.
\end{abstract}

\maketitle

\section{Introduction} Understanding quantum chaos and its relation to the thermalization process is one of the greatest challenges of quantum statistical physics. Traditionally, study of quantum chaos has been limited to statistics of energy level spacings in combination with a series of semiclassical methods. In the past few years, study of four-point out-of-time-ordered correlators (OTOCs)\cite{larkin,Kitaev,Maldacena} as a signature of quantum chaos has attracted a surge of theoretical and experimental interest. The exponential growth rate of OTOCs, i.e. scrambling rate ($\lambda_L$) generalizes the notion of Lyapunov exponent from classical physics to quantum chaos\footnote{Though the relation between $\lambda_L$ and the classical Lyapunov exponent is subtle and not straightforward (see Ref.\onlinecite{Efim1}).}.

OTOCs were first introduced in the context of quasi-classical methods in superconductivity\cite{larkin}. More recently Refs.~\onlinecite{Kitaev,Maldacena} revived OTOCs by discovering a fundamental bound on $\lambda_L$. Following these seminal works, OTOCs have been studied in a plethora of many-body quantum systems\cite{otoc1,otoc2,otoc3,otoc4,otoc5,otoc6,otoc7,otoc8,otoc9,otoc10,otoc11,otoc12,otoc13,otoc14,otoc15,otoc16,otoc17,otoc18}. On the experimental side, a series of proposals on how to measure OTOCs\cite{prop1,prop2,prop3,prop4,prop5,prop6,prop7,prop8,prop9} as well as some preliminary measurements\cite{exp1,exp2,exp3,exp4} have already been reported.

 The main motivation of the present work is to study the scrambling rate $\lambda_L$ in the iconic example of the Dicke model\cite{dicke,dmrev} (DM). DM describes a zero dimensional (no spatial structure) collection of $N$ spin $1/2$ degrees of freedom (e.g. two level atoms) interacting with a single Bosonic mode. Above some critical value of coupling $g=g_c$, the DM undergoes a phase transition to a super-radiant phase that is characterized by a nonzero mean displacement of the Bosonic field\cite{dc1,dc2}. DM hosts a series of quantum and classical signatures of chaos that are particularly strong in the super-radiant phase\cite{dc-1,dc0,dc1,dc2,dc3,dc4,dc5,dc6,dc7,dc8}. In addition to theoretical interest, experimental platforms to measure OTOCs in the DM already exist\cite{exp0,exp1}. These features make DM a particularly good candidate to study quantum signatures of chaos.

 In this article, we use the Majorana Fermion representation of spin $1/2$ systems\cite{mj1,mj2,mj3,dmj1,dmj2} in combination with the diagrammatic method of Ref.~\onlinecite{Stanford} to compute $\lambda_L$ throughout the phase diagram to leading order in $1/N$. We take the system to be in thermal equilibrium and study OTOCs associated with two different operators. Our formalism does not rely on the assumption of total spin conservation which is not justified in most experimental realizations. We show that the dominant terms contributing to the scrambling rate are given by two \textit{different} sets of diagrams in the normal and super-radiant phases respectively. We find that at low temperatures, the appearance of chaotic behavior is limited to (a) close proximity of the critical point and (b) the super-radiant phase. Whereas, in high temperatures $\lambda_L\neq 0$ in both the normal and super-radiant phases. We provide an example of how the scrambling rates associated with two different operators can be different. We provide a discussion of our results in relation with previous semiclassical studies of chaos in the DM.  In the process we also derive a new effective theory for the super-radiant phase at finite temperature.

 The rest of this paper is organized as follows: in Section \ref{Model} we introduce the DM Hamiltonian and the OTOCs we are interested in. Section \ref{majorana} presents the Majorana Fermion representation of spin $1/2$ systems. In Sections \ref{normal} and \ref{sec:Effective theory in the super-radiant phase} we use the Majorana operators to obtain effective theories in the normal and super-radiant phases respectively. In Section \ref{cotoc} we review the diagrammatic method used to compute the scrambling rate. Section \ref{cotoc1} contains explicit diagrammatic calculations used to compute $\lambda_L$ in the DM. Our final results are stated and discussed in Section \ref{reslt}. We end with a brief summary and conclusion in Section \ref{d&c}.

\section{Model}\label{Model}  The Hamiltonian describing DM is given by,
\begin{align}\label{hs}
H=\omega_0 a^\dagger a + \omega_z \sum_{j=1}^N \sigma_j^z + \frac{2g}{\sqrt{N}} \sum_{j=1}^N\sigma_j^x (a+a^\dagger).
\end{align}
Here $\sigma$'s correspond to the usual spin $1/2$ operators and $a,a^\dagger$ are the standard Bosonic annihilation and creation operators.

We also define the real Bosonic field $\phi$  (``position" degree of freedom of the harmonic oscillator) as,
\begin{align}
\phi=a+a^{\dagger}.
\end{align}
 The total spin $S_{tot}^2=(\Sigma_i \sigma_i^x)^2+(\Sigma_i \sigma^y_i)^2+(\Sigma_i \sigma^z_i)^2$ is conserved. Furthermore, the Hamiltonian is invariant under a parity transformation,
\begin{equation}\label{prt}
  \Pi=e^{i\pi (a^\dagger a+S_z)},
\end{equation}
which rotates the spins around the $z$ axis by $\pi$ and takes $\phi$ to $-\phi$.

At zero temperature and in the large $N$ limit ($N\rightarrow \infty$), it can be shown that at a critical value of coupling $g_c=\sqrt{\omega_0\omega_z}/2$, this model undergoes a phase transition from the normal phase ($\expval{a}= 0$) at $g<g_c$ to a super-radiant phase ($\expval{a}\neq 0$) at $g>g_c$. The parity symmetry described in Eq.\ref{prt} is  spontaneously broken in the super-radiant phase.

In this work, we are interested in the following OTOCs,
\begin{align}\label{c1}
&C_{\sigma_z}(t)=-\frac{1}{N^2}\sum_{j,k=1}^N\expval{ [\sigma_j^z(t),\sigma_k^z]^2}_\beta,\nonumber   \\ &C_{\phi}(t)=-\expval{ [\phi(t),\phi]^2}_\beta.
\end{align}
However, for the calculations in this paper, it is more convenient to work with a ``regulated" form of OTOCs,
\begin{align}\label{c11}
&\mathcal{C}_{\sigma_z}(t)=-\frac{1}{N^2}\sum_{j,k=1}^N\Tr{ \sqrt{\rho}[\sigma_j^z(t),\sigma_k^z] \sqrt{\rho}[\sigma_j^z(t),\sigma_k^z] },\nonumber   \\
&\mathcal{C}_{\phi}(t)=-\Tr{ \sqrt{\rho}[\phi(t),\phi]\sqrt{\rho}[\phi(t),\phi]},
\end{align}
 where $\rho$ is the thermal density matrix. In a chaotic system the early time behavior of $\mathcal{C}(t)$ is expected to be proportional to $\propto e^{\lambda_L t}$, where $\lambda_L$ is the scrambling rate. We remark that according to a recent study\cite{Galitski} the early time behavior of $C(t)$ and $\mathcal{C} (t)$ can be different, and that the scrambling rate associated with  $\mathcal{C} (t)$ is the true measure of may-body chaos in quantum systems.

  Single spin operators $\sigma_j^z$ do not commute with the total spin operator $S_{tot}^2$ and therefore, usual methods applied to DM that rely on total spin conservation are of limited use in calculating $\mathcal{C}_{\sigma_z}(t)$.

 A powerful diagrammatic method to compute $\lambda_L$ has been described in Ref.\onlinecite{Stanford}. To apply the formalism of Ref.\onlinecite{Stanford} to our problem, we need to switch from using spin operators $\sigma_j$ to a form that is amenable to Wick's theorem (and therefore perturbation theory). Refs.\onlinecite{dmj1,dmj2} showed that a straightforward way to do this is to use the Majorana Fermion representation of the spin $1/2$ systems.

  Before proceeding further, we'd like to clarify that  we will not attempt to calculate $\mathcal{C}_{\sigma_z}(t)$ directly. Instead, we focus on the OTOC associated with a closely related quantity ($\tilde{\sigma}_z$ defined in Section.\ref{sec:Effective theory in the super-radiant phase}) that reduces to $\sigma_z$ in the normal phase.

\section{Majorana Fermion representation of the spin $1/2$}\label{majorana}
Following Refs.\onlinecite{mj1,mj2,mj3} let us consider the following Majorana Fermion representation the spin $1/2$ operators,
\begin{align}\label{def}
&\sigma_j^x = \frac{1}{2}\eta_j (f_j-f_j^\dagger) \nonumber \\
&\sigma_j^y = \frac{i}{2}\eta_j (f_j+f_j^\dagger) \nonumber \\
&\sigma_j^z = f_j^\dagger f_j -1/2,
\end{align}
where $f_j,f_j^\dagger$ represent Fermionic creation and annihilation operators which satisfy $\{ f_j,f_k^\dagger \}=\delta_{jk}$. $\eta_j$ represents a Majorana Fermion obeying $\{ \eta_j ,\eta_k  \}=2 \delta_{jk}$. The redundancy of this Fermion representation leads to a "$\mathbb{Z}_2$ gauge" symmetry. The $\mathbb{Z}_2$ gauge symmetry generator $\gamma_j$ is given by,
\begin{align}
\gamma_j=(2 f_j^\dagger f_j -1)\eta_j.
\end{align}
The operator $\gamma_j$ commutes with all other spin operators $\sigma_k^x,\sigma_k^y,\sigma_k^z$ and is therefore a constant of motion $\gamma_j (t)=\gamma_j (0)$. Note that $\gamma_j^2=1$. It is also useful to realize that,
\begin{align}\label{mr}
\sigma^+_j= f_j \gamma_j \quad;\quad \sigma_j^z=\frac{1}{2} \eta_j \gamma_j
\end{align}

Using Eq.\eqref{mr}, we can rewrite Eq.\eqref{c1},\eqref{c11} as,
\begin{align}\label{regulatedOTOC}
&C_{\sigma_z}(t)=\frac{1}{16 N^2}\sum_{j,k=1}^N \langle \{\eta_j(t),\eta_k\}^2\rangle_\beta\nonumber \\
&\mathcal{C}_{\sigma_z}(t)=\frac{1}{16 N^2}\sum_{j,k=1}^N \Tr{  \sqrt{\rho} \{\eta_j(t),\eta_k\} \sqrt{\rho} \{\eta_j(t),\eta_k\}}.
\end{align}
This simple form is a direct consequence of $\gamma$ being a constant of motion. If we had naively used Eq.\eqref{def} to write $C_{\sigma_z}(t)$, we would have ended up with complicated eight point correlation functions.

\section{Theory in the normal phase}\label{normal}

In this section, we follow the approach of Ref.\onlinecite{dmj1} to describe the theory in the normal phase.

\begin{figure}[t]
\centering
\includegraphics[width=\columnwidth,keepaspectratio]{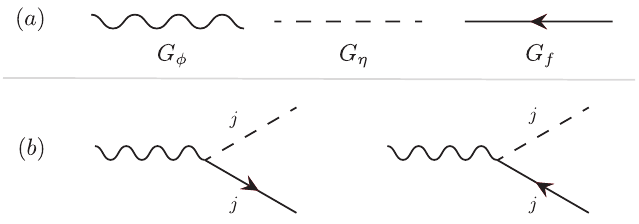}
\caption{(a) Bare Green's functions of Boson, Majorana and Fermionic fields (b) interaction vertices in the normal phase \label{fig:1.pdf}}
\end{figure}

Using Eq.\eqref{def} we can rewrite the DM Hamiltonian (Eq.\eqref{hs}) in terms of the Fermionic variables (up to a constant),
\begin{align}\label{hn}
H=\omega_0 a^\dagger a + \omega_z \sum_{j=1}^N f^\dagger_j f_j + \frac{g}{\sqrt{N}} \sum_{j=1}^N \eta_j (f_j-f_j^\dagger) (a+a^\dagger).
\end{align}
 Advantage of this form is that it allows for a systematic large $N$ diagrammatic treatment, in which we still have access to individual spin operators. The first two terms describe a free quadratic theory and the last term describes interaction vertices shown in Fig.\ref{fig:1.pdf}(b).

As mentioned before, we use the real Bosonic field $\phi=a+a^\dagger,$ instead of $a,a^\dagger$. Bare Matsubara Green's functions can now be written in the usual form,
\begin{align}
G^0_{\phi}(i \omega_n&)=\frac{2\omega_0}{(i \omega_n)^2+\omega_0^2}, \quad G^0_{\eta}(i \omega_n)=\frac{2}{i \omega_n}, \nonumber \\
G^0_{f}(i \omega_n&)=\frac{1}{i \omega_n-\omega_z}, \quad G^0_{f^\dagger}(i \omega_n)=\frac{1}{i \omega_n+\omega_z}.
\end{align}
Diagrammatic representation of Green's functions and interaction vertices is shown in Fig.\ref{fig:1.pdf}.

We then use Eq.\eqref{hn} to calculate self energies associated with the fields $\eta,f,\phi$. To leading order in $1/N$ the Bosonic self energy is given by,
\begin{align}\label{seb}
&\Sigma_\phi (i \omega_n)=\vcenter{\hbox{\includegraphics[width=0.15\columnwidth,keepaspectratio]{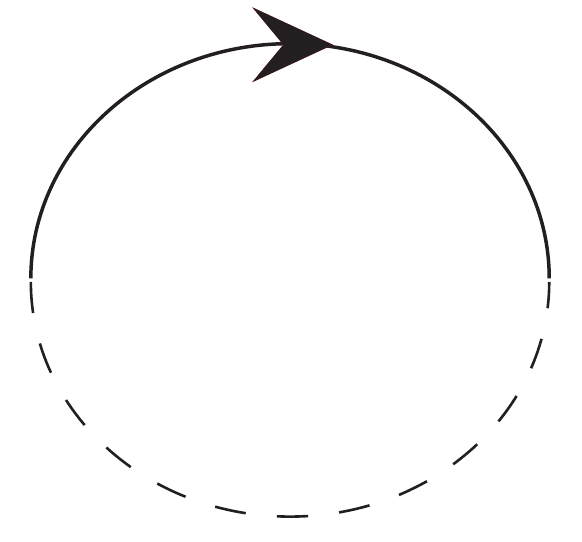}}}+ \vcenter{\hbox{\includegraphics[width=0.15\columnwidth,keepaspectratio]{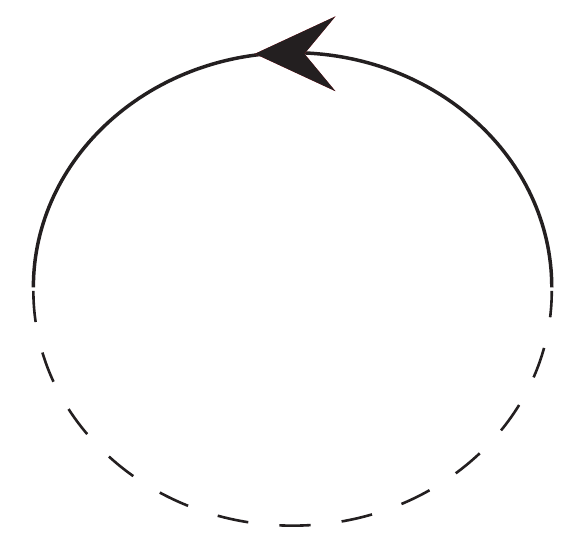}}}+\mathcal{O}(1/N)\nonumber \\
&=-\frac{2g^2\omega_z}{\omega_n^2+\omega_z^2}\tanh(\beta \omega_z/2)+\mathcal{O}(1/N).
\end{align}
In the diagrams above, an implicit sum over Majorana and Fermion fields' index $j$ has been assumed. The sum over the internal index cancels the factor of $1/N$ arising form the two vertices. It is easy to see that Majorana and Fermion self energies are both zero to zeroth order in $1/N$, i.e. $\Sigma_\eta (i \omega_n)=\Sigma_f (i \omega_n)=\mathcal{O}(1/N)$. The simple form of this equations is what makes the model exactly solvable in the large $N$ limit.

Using the self energy expression in Eq.\eqref{seb}, we write the dressed Bosonic propagator (to leading order in $1/N$) as,
\begin{align}
G_\phi(i \omega_n) &=\frac{1}{G_\phi^{0^{-1}}(i\omega_n)-\Sigma_\phi (i \omega_n)}\nonumber \\
&=\frac{2\omega_0((i \omega_n)^2-\omega_z^2)}{((i \omega_n)^2-\omega_+^2)((i \omega_n)^2-\omega_-^2)},\end{align}
where $\omega_\pm$ are given by,
\begin{align}\label{wpm}
&\omega_{\pm}^2=\frac{\omega_0^2+\omega_z^2}{2}\nonumber \\
&\pm \sqrt{(\frac{\omega_0^2+\omega_z^2}{2})^2-\omega_0^2\omega_z^2+4g^2\omega_0\omega_z\tanh(\beta\omega_z/2)}.
\end{align}
Note that at \textit{zero temperature} limit of these results are the same as the spectrum derived using the Holstein-Primakoff Representation\cite{dc1,dc2}.

This expression signals a finite temperature phase transition (divergence of $G_\phi(0)$) at,
\begin{align}\label{gc}
g_c=\frac{1}{2}\sqrt{\frac{\omega_0\omega_z}{\tanh(\beta \omega_z /2)}}.
\end{align}
 At couplings $g>g_c$ one of the poles becomes ``positive imaginary", which indicates an instability of the perturbation theory. As we'll show below this can be remedied by assuming a nonzero expectation value for the Bosonic field, i.e. $\langle \phi \rangle \ne 0$.

It is worth noting that all Green's functions' poles are \textit{real} to zeroth order in $1/N$. In-order to obtain the leading order correction to the imaginary part of the poles (i.e. relaxation time), one needs to consider two loop diagrams. In this work we ignore such corrections and leave their calculation to future work.

\section{Effective theory in the super-radiant phase}\label{sec:Effective theory in the super-radiant phase}

The breakdown of perturbation theory for $g>g_c$ is related to the fact that in the strong interaction limit, the Bosonic field acquires non-zero macroscopic vacuum expectation value,
\begin{equation}
  \expval{a} \sim \sqrt{N}.
\end{equation}
It can also be understood as the displacement of the action's saddle point in the path integral description of the theory; consequently, the original Bosonic and Fermionic fields are no longer suitable degrees of freedom to describe the low energy physics of the system.

 To obtain the appropriate fields in the super-radiant phase, we start by defining the new field operator $\tilde a$ as,
\begin{equation}\label{atld}
  \tilde a = a - \frac{\alpha}{2}\sqrt{N}
\end{equation}
for some constant real $\alpha$. Our goal is to find the value of $\alpha$ such that the vacuum expectation value of $\tilde \phi = \tilde a + \tilde a^\dagger$ field becomes zero.

\begin{figure}
  \includegraphics[width=0.25\textwidth]{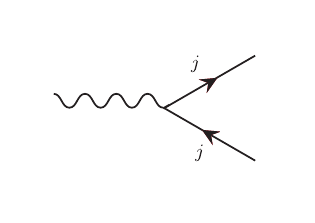}
  \caption{The new interaction vertex in the super-radiant phase.}
  \label{nvtx}
\end{figure}

If we rewrite Hamiltonian \eqref{hs} in terms of $\tilde a$ we get (up to a constant),
\begin{align}\label{h_shifted}
  H=&\omega_0 \tilde a^\dagger \tilde a-\frac{\alpha \omega_0 \sqrt{N}}{2} (\tilde a^\dagger + \tilde a)+\sum_{j=1}^N\qty[\omega_z \sigma^z_j - 2 g \alpha \sigma^x_j]\nonumber \\
      &+\frac{2 g}{\sqrt{N}}\sum_{j=1}^N\sigma^x_j(\tilde a^\dagger+\tilde a).
\end{align}
The form of Hamiltonian \eqref{h_shifted} suggests defining a new set of \textit{rotated} spin operators,
\begin{align}
&\tilde{\sigma}_z=\cos (\theta) \sigma_z + \sin (\theta) \sigma_x \nonumber \\
&\tilde{\sigma}_x= -\sin (\theta) \sigma_z + \cos (\theta) \sigma_x,
\end{align}
with the angle $\theta$ defined as,
\begin{equation}\label{thetae}
\sin(\theta)\equiv\frac{-2\alpha g}{\tilde \omega_z},
\end{equation}
where $$ \tilde \omega_z=\sqrt{\omega_z^2+4 g^2 \alpha ^2}.$$
Using these new variables, Hamiltonian \eqref{h_shifted} can be written as,
\begin{align}
    H=& \omega_0 \tilde a^\dagger \tilde a+\tilde \omega_z\sum_{j=1}^N\tilde \sigma^z_j+\frac{2 g \cos(\theta)}{\sqrt{N}}\sum_{j=1}^N\tilde \sigma^x_j(\tilde a^\dagger+\tilde a)\nonumber \\
      &+\frac{2 g \sin(\theta)}{\sqrt{N}}\sum_{j=1}^N\tilde \sigma^z_j(\tilde a^\dagger+\tilde a)-\frac{\alpha \omega_0 \sqrt{N}}{2} (\tilde a^\dagger + \tilde a).
\end{align}
Finally, we use Majorana representation in this new rotated frame to exchange spin operators for Majorana Fermions,
\begin{align}\label{ef2}
  H=& \omega_0 \tilde a^\dagger \tilde a+\tilde \omega_z\sum_{j=1}^N\tilde f^\dagger_j \tilde f_j\nonumber \\
  &+\frac{g \cos(\theta)}{\sqrt{N}}\sum_{j=1}^N\tilde \eta_j(\tilde f_j - \tilde f_j^\dagger)(\tilde a^\dagger+\tilde a)\nonumber\\
  &+\frac{2 g \sin(\theta)}{\sqrt{N}}\sum_{j=1}^N\tilde f_j^\dagger \tilde f_j (\tilde a^\dagger+\tilde a)\nonumber \\
  &-\frac{\sqrt{N}}{2}[\alpha \omega_0 + 2 g \sin (\theta)] (\tilde a^\dagger + \tilde a),
\end{align}
where $\tilde f_j$ and $\tilde \eta_j$ are related to $\tilde \sigma_x$ and $\tilde \sigma_z$ operators according to Eq.\eqref{def}. Note that in this basis, we have an additional interaction vertex shown in Fig.~\ref{nvtx}. We emphasize that presence of this new interaction vertex is the main feature distinguishing normal and super-radiant phases. Crucially, this term breaks the parity symmetry associated with the normal phase (Eq.\eqref{prt}).

 We assume for a given coupling constant $g$ and temperature $T$, the value of $\alpha$ is chosen such that $\expval{\tilde\phi}=0$. Then we can use diagrammatic method to solve for the value of $\alpha$ self consistently. In the large $N$ limit, the leading order contribution to $\expval{\tilde \phi}$ is given by the following diagrams,
 \begin{align}\label{expphi}
   \expval{\tilde \phi}&= \vcenter{\hbox{\includegraphics{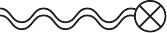}}} + \vcenter{\hbox{\includegraphics{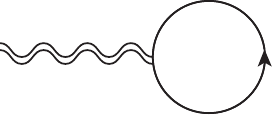}}} \nonumber\\
                      &=\vcenter{\hbox{\includegraphics{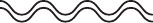}}} \times \Big(\vcenter{\hbox{\includegraphics{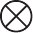}}} + \vcenter{\hbox{\includegraphics{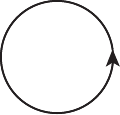}}}\Big)
 \end{align}

The first diagram comes from the $(\tilde a + \tilde a^\dagger)$ term in the Hamiltonian and the second term is related to the $\tilde f_j^\dagger \tilde f_j (\tilde a^\dagger+\tilde a)$ interaction term. The double wavy line represents the dressed $\tilde \phi$ field propagator\footnote{Since the Boson propagator is renormalized by the interaction even in the limit $N\to \infty$, we used the dressed propagator here. However its exact form isn't important for the purpose of current calculations.} and the solid line represents the Fermionic propagator. All other contributions to $\expval{\tilde \phi}$ are of sub-leading order in $1/N$. To satisfy $\expval{\tilde \phi}=0$, we demand the expression inside parentheses in Eq.\eqref{expphi} to vanish,
\begin{equation}
  -\frac{2 g \sin \theta}{\sqrt{N}}\sum_j \expval{\tilde f_j^\dagger \tilde f_j} + \frac{\sqrt{N}}{2}\qty[\alpha \omega_0 + 2  \sin \theta]=0.
\end{equation}
To leading order in $1/N$, we can replace $\expval{\tilde{f}_j^\dagger \tilde{f}_j}$ with Dirac distribution function and re-arrange the terms to arrive at the following equation for $\alpha$,
\begin{equation}\label{alpheq}
  \alpha\qty[g^2-\frac{\omega_0 \tilde \omega_z }{4 \tanh(\beta \tilde \omega_z/2)}]=0.
\end{equation}
Note that $\alpha=0$ always satisfies this equation. This solution corresponds to the original fields we used to describe the normal phase. For $g>g_c$ the expression in the brackets also has two real roots with the same magnitude and the opposite signs. The corresponding solutions are related to each other by the parity operator defined in Eq. \eqref{prt}. Note that according to Eq.\eqref{atld}, root of Eq.\eqref{alpheq} corresponds to the vacuum expectation value of the original Bosonic field $\phi$,
\begin{equation}
  \expval{\phi}=\alpha\sqrt{N}.
\end{equation}
As shown in the previous section, $\alpha=0$ solution is unstable in the super radiant phase and the system chooses one of the other non-zero roots and hence spontaneously breaks the parity symmetry.

\begin{figure}[!t]
  \includegraphics[width=\columnwidth]{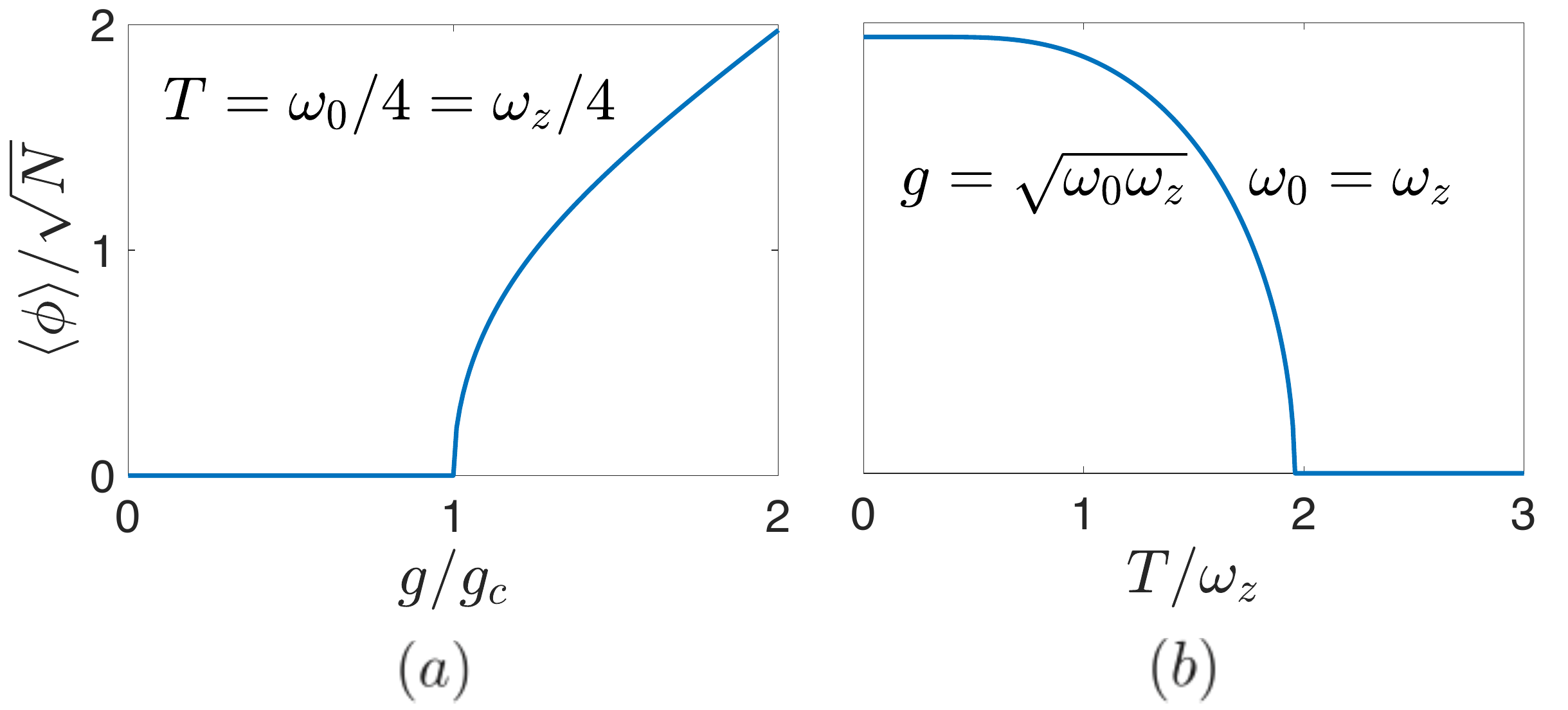}
  \caption{(a) $\expval{\phi}/\sqrt{N}=\alpha$ versus $g$ for fixed value of $T$. (b) $\expval{\phi}/\sqrt{N}$ versus $T$ for fixed value of $g$.}
  \label{alpha}
\end{figure}

The value of $\expval{\phi}/\sqrt{N}$ versus $g$ at fixed temperature $T=\omega_z/4=\omega_0/4$ is plotted in Fig. \ref{alpha}(a). The horizontal axis is $g/g_c$ where $g_c$ is the critical value of $g$ at temperature $T$ as given in Eq.\eqref{gc}. As expected, for $g<g_c$ the system is in the normal phase and $\expval{\phi}=0$ whereas for $g>g_c$, $\expval{\phi}$ becomes non-zero and grows as one further increases the interaction strength $g$.

In Fig. \ref{alpha}(b) we also look at $\expval{\phi}/\sqrt{N}$ versus temperature for a fixed value of coupling constant $g_0=\sqrt{\omega_0 \omega_z}$. Note that by increasing the temperature, the system will eventually go back to the normal phase. The critical temperature for a given fixed $g$ can be calculated by inverting Eq.\eqref{gc} to solve for $T_c$,
\begin{equation}
  T_c=\frac{\omega_z}{2\tanh^{-1}(\frac{\omega_z \omega_0}{4 g^2})}.
\end{equation}

 This particular form of $\alpha (g,T)$ in combination with Eqs.\eqref{thetae},\eqref{ef2} defines the effective theory in the super-radiant phase. This theory also applies to the normal phase by setting $\alpha=\theta=0$ and hence, from now on we use this theory in the entire phase diagram. To the best of our knowledge this effective theory as well as the average value of $\expval{a}=\sqrt{N}\alpha (g,T)/2$ at nonzero temperatures (plotted in Fig.\ref{alpha}) were not known before.

 Since $\alpha$ is a function of temperature, the parameters of Hamiltonian \eqref{ef2} become temperature dependent. Note that both $\tilde \omega_z$ and $\theta$ are functions of $\alpha$ and hence functions of $g$ and $T$.

We remark that Eq.\eqref{ef2} implies that the natural variables describing the system are $\tilde{\sigma}_z,\tilde\phi$. These variables reduce to the original $\sigma_z,\phi$ in the normal phase, whereas in the super-radiant phase, they are related to $\sigma_z,\phi$ via \textit{rotation} and \textit{translation} respectively.

Green's functions of the theory in the super-radiant phase can now be calculated using diagrammatic techniques. Note that by setting Eq.\eqref{expphi} to zero, we have insured that the terms associated with $(a+a^\dagger)$ and tadpole diagrams always cancel each other, i.e. neither one needs to be included in any diagram.

Bare Green's functions have the same form as in the normal phase, only with new parameters,
\begin{align}
G^0_{\tilde \phi}(i \omega_n&)=\frac{2\omega_0}{(i \omega_n)^2+\omega_0^2}, \quad G^0_{\tilde \eta}(i \omega_n)=\frac{2}{i \omega_n}, \nonumber \\
G^0_{\tilde f}(i \omega_n&)=\frac{1}{i \omega_n-\tilde \omega_z}, \quad G^0_{\tilde{f}^\dagger}(i \omega_n)=\frac{1}{i \omega_n+\tilde \omega_z}. \nonumber \\
\end{align}
Similar to the normal phase, self energies associated with $\tilde \eta$ and $\tilde f$ fields are of the order $1/N$ and vanish in the large $N$ limit. However, the Boson's self energy has an additional contribution from $\tilde f^\dagger \tilde f \tilde \phi$ vertex,
\begin{align}\label{seb-sr}
\Sigma_{\tilde \phi} (i \omega_n)=&\vcenter{\hbox{\includegraphics[width=0.15\columnwidth,keepaspectratio]{e1.pdf}}}+ \vcenter{\hbox{\includegraphics[width=0.15\columnwidth,keepaspectratio]{e2.pdf}}}+\vcenter{\hbox{\includegraphics[width=0.15\columnwidth,keepaspectratio]{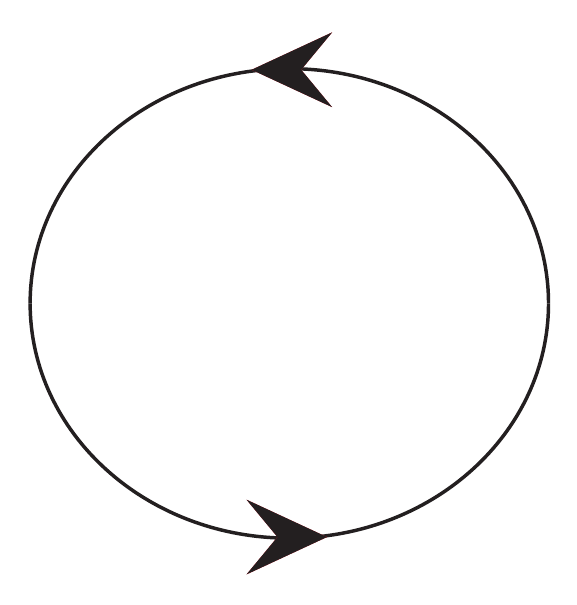}}}+\mathcal{O}(1/N)\nonumber \\
=&-\frac{2g^2\cos^2\theta\tilde \omega_z}{\omega_n^2+\tilde \omega_z^2}\tanh(\beta \tilde \omega_z/2)\nonumber\\
  &+4g^2\sin^2\theta \, n_F'(\tilde \omega_z)\, \delta_{\omega_n,0},
\end{align}
where $n_F'$ is the derivative of the Fermi function. The $\delta_{\omega_n,0}$ term only adds a \textit{time independent constant} to the imaginary time Green's function. This constant can be absorbed in the definition of $\tilde \phi$ field and therefore does not affect the retarded Green's function. As we'll show in the next section, for computing OTOCs we only need the retarded Green's functions. To the zeroth order in $1/N$, we can write the dressed retarded Bosonic propagator as,
\begin{equation}
  G^R_{\tilde \phi}(\omega) =\frac{2\omega_0(\omega^2-\tilde \omega_z^2)}{((\omega+i\varepsilon)^2-\tilde \omega_+^2)((\omega+i\varepsilon)^2-\tilde\omega_-^2)},
\end{equation}
where $\tilde \omega_\pm$ are given by the same expression as in Eq.\eqref{wpm}, but with $\omega_z$ replaced by $\tilde \omega_z$ and $g$ replaced by $g\cos\theta$. Similar to the normal phase, these results reproduce the spectrum derived using the Holstein-Primakoff representation in the super-radiant phase\cite{dc1,dc2}.

Analogous to the normal phase, imaginary part of the Green's functions' poles are of sub-leading order in $1/N$  and involve two loop diagrams. These corrections are ignored here.

As mentioned earlier the natural variables describing the system are $\tilde{\sigma}_z,\tilde\phi$. Motivated by this observation, we study the scrambling rates associated with $\tilde{\sigma}_z,\tilde\phi$, i.e., $\mathcal{C}_{\tilde\sigma_z}(t),\mathcal{C}_{\tilde\phi} (t)$ (defined similar to Eq.\eqref{c11}). However, note that $\mathcal{C}_{\tilde\phi} (t)=\mathcal{C}_{\phi} (t)$, whereas $\mathcal{C}_{\tilde\sigma_z}(t)$ is equivalent to $\mathcal{C}_{\sigma_z}(t)$ only in the normal phase.

\section{Diagrammatic rules for calculating OTOC}\label{cotoc}
\begin{figure*}[t]
\centering
\includegraphics[width=2\columnwidth,keepaspectratio]{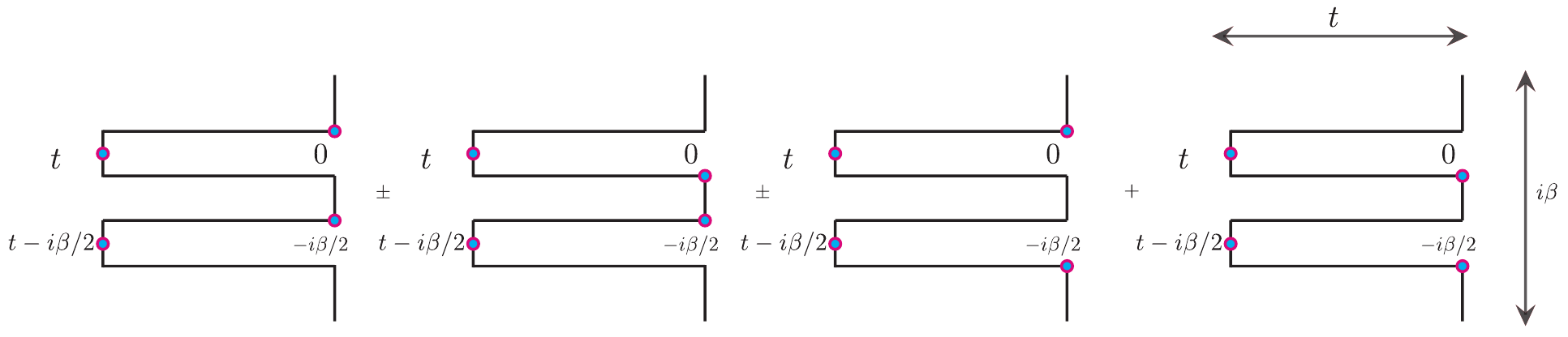}
\caption{Contour $c$ used for evaluating OTOCs. Horizontal and vertical lines are real and imaginary time axes respectively. Circles on the contour represent field operators and their ordering. Plus and minus signs correspond to anti-commutator and commutator respectively (e.g. $\mathcal{C}_{\sigma_z}/\mathcal{C}_{\phi}$). \label{fig:2.pdf}}
\end{figure*}

\begin{figure*}[ht]
\centering
\includegraphics[width=2\columnwidth,keepaspectratio]{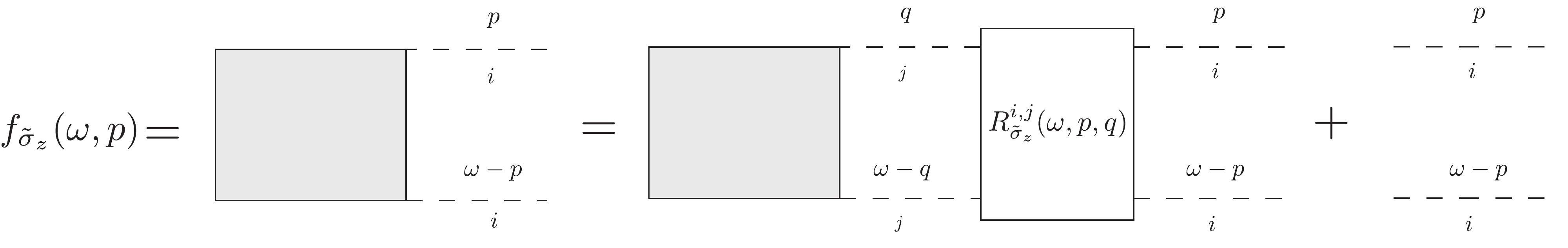}
\caption{A diagrammatic equation for $f_{\tilde\sigma_z}(\omega,p)$. Internal indices $i,j$ are summed over.\label{fig:3.pdf}}
\end{figure*}
Since OTOCs are not time ordered, calculating them using usual methods of quantum field theory is difficult. In this section we review the method developed in Ref.\onlinecite{Stanford} to calculate OTOCs.

We start by rewriting the ``regulated" OTOCs in the following form,
\begin{align}
 &\mathcal{C} _{\tilde\sigma_z}(t)= \frac{1}{N^2}\sum_{j,k=1}^N\expval{\{\tilde\eta_j(t-i\beta /2),\tilde\eta_k(-i\beta /2)\}   \{\tilde\eta_j(t),\tilde\eta_k\}  } \nonumber \\
&\mathcal{C}_{\tilde\phi}(t)=-\expval{  [\tilde\phi(t-i \beta/2),\tilde\phi(-i\beta/2)]  [\tilde\phi(t),\tilde\phi] }.
\end{align}
 Operators in this new form are now ordered along a contour $c$ that goes through both real and imaginary times (Fig.\ref{fig:2.pdf}). We then switch to the interaction picture and expand $\mathcal{C}$ in powers of the interaction vertex to arrive at a set of diagrammatic rules for calculating OTOC.

 Before stating the rules of diagrammatic calculation, we need to introduce ``Wightman functions" that correspond to propagators along the thermal circle,
 \begin{align}
 &G^W_{\tilde\phi}(t)=\expval{\tilde\phi (t-i\beta/2)\tilde\phi (0)}\nonumber\\
 &G^W_{\tilde\eta}(t)=\expval{\tilde\eta (t-i\beta/2)\tilde\eta (0)} \nonumber \\
 &G^W_{\tilde{f}}(t)=\expval{\tilde{f} (t-i\beta/2)\tilde{f}^\dagger (0)}\nonumber\\
 &G^W_{\tilde{f}^\dagger}(t)=\expval{\tilde{f}^\dagger (t-i\beta/2)\tilde{f} (0)}.
 \end{align}
 We need the explicit form of Fermionic Wightman functions in frequency space (to leading order in $1/N$),
\begin{align}\label{wmf}
&G^W_{\tilde\eta}(\omega)=2\pi \delta(\omega) \nonumber \\
&G^W_{\tilde{f}}(\omega)=\frac{2\pi \delta(\omega-\tilde\omega_z)}{2\cosh(\beta \tilde\omega_z/2)}\nonumber \\
&G^W_{\tilde{f}^\dagger}(\omega)=\frac{2\pi \delta(\omega+\tilde\omega_z)}{2\cosh(\beta \tilde\omega_z/2)}.
 \end{align}

 Rules of diagrammatic calculation can now be summarized as follows (for a detailed derivation look at Refs.\onlinecite{Stanford,otoc2}):

 1. Horizontal direction represents the real time and correspondingly horizontal lines correspond to \textit{dressed} retarded Green's functions $i G^R$ (self energy diagrams should not be included here). Vertical direction represents the imaginary time and correspondingly non-horizontal (vertical and crossed) lines correspond to Wightman propagators $G^W$.

 2. Vertices are only added along the real time folds. Vertex insertions along the imaginary part of the contour will dress the thermal density matrix (from $\rho_0=\frac{\exp(-\beta H_0)}{\mathcal{Z}}$ of free theory to the $\rho=\frac{\exp(-\beta H)}{\mathcal{Z}}$ of interacting theory). However the growth rate of OTOCs is expected to be independent of the exact form of the thermal state\cite{Stanford,otoc1,otoc2,otoc6,otoc16}.

 The total sign associated with Wick contractions should be accounted for in each diagram.

\section{Diagrammatic calculation of the OTOC}\label{cotoc1}
In this section we use the diagrammatic method to obtain explicit integral equations for $\mathcal{C}_{\tilde\sigma_z}(t)$ and $\mathcal{C}_{\tilde\phi}(t)$. In the next section we use these equations to obtain the associated scrambling rates $\lambda_L^{\tilde\phi}$ and $\lambda_L^{\tilde{\sigma_z}}$.
\subsection{Diagrammatic form of $\mathcal{C}_{\tilde\sigma_z}(t)$}\label{csz}

\begin{figure*}[!t]
  \includegraphics[width=\textwidth]{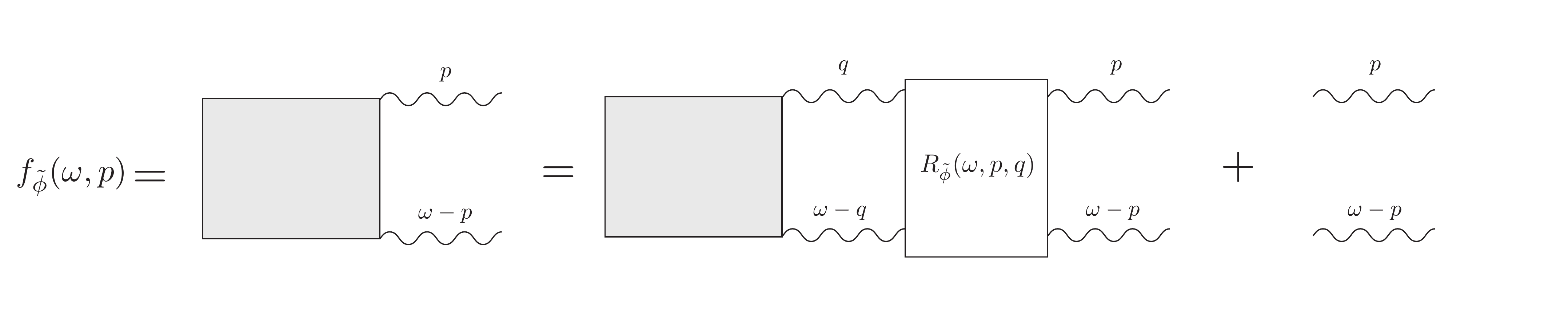}
  \caption{A diagrammatic equation for $f_{\tilde\phi}(\omega,p)$.}
  \label{b-s}
\end{figure*}

\begin{figure*}[!t]
  \includegraphics[width=\textwidth]{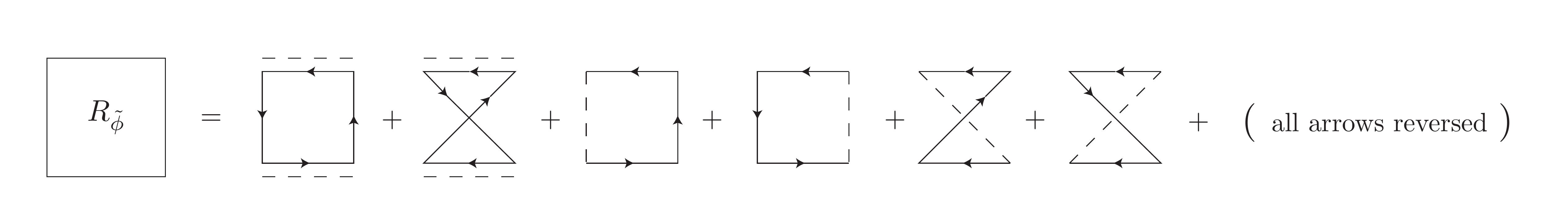}
  \caption{Bosonic rung function $R_{\tilde\phi}(\omega,p,q)$ to leading order in $1/N$. An implicit sum over the internal index is assumed. Double horizontal lines in the first two terms correspond to the sum of Fermion and Majorana propagators (the first two terms correspond to a total of $16$ diagrams).}
  \label{Boson_rung.pdf}
\end{figure*}
We begin by defining $f_{\tilde\sigma_z}(\omega,p)$ as,
\begin{align}
\mathcal{C}_{\tilde\sigma_z}(\omega)=\frac{1}{N^2}\int_{-\infty}^\infty \frac{dp}{2\pi} f_{\tilde\sigma_z}(\omega,p).
\end{align}
$f_{\tilde\sigma_z}(\omega,p)$  is comprised of a series of diagrams with a pair of Majorana propagators attached to both the right and left ends of each diagram. This set of diagrams can be summed over using a Bethe-Saltpeter type equation. A diagrammatic equation for $f_{\tilde\sigma_z}(\omega,p)$ is shown in Fig.\ref{fig:3.pdf}. This equation can be explicitly written as,
\begin{align}
&f_{\tilde\sigma_z}(\omega,p)= - G^R_{\tilde\eta} (p) G^R_{\tilde\eta}(\omega-p) \nonumber \\
&\times \Big[N+ \sum_{i,j} \int \frac{dq}{2\pi}  R^{i,j}_{\tilde\sigma_z}(\omega,p,q) (\frac{1}{N} f_{\tilde\sigma_z}(\omega,p))\Big].
\end{align}
As in Ref.~\onlinecite{Stanford}, we notice that the first term in the square bracket does not give rise to exponential growth. This term can then be dropped for the purpose of calculating $\lambda_L$,
\begin{align}\label{m1}
&f_{\tilde\sigma_z}(\omega,p)= - G^R_{\tilde\eta} (p) G^R_{\tilde\eta}(\omega-p) \nonumber \\
&\times  \frac{1}{N}\sum_{i,j} \int \frac{dq}{2\pi}  R^{i,j}_{\tilde\sigma_z}(\omega,p,q)  f_{\tilde\sigma_z}(\omega,p).
\end{align}

To leading order in $1/N$ the rung function $R^{i,j}_{\tilde\sigma_z}(\omega,p,q)$ can be approximated by a single diagram,
\begin{align}\label{rmj}
&R^{i,j}_{\tilde\sigma_z}(\omega,p,q)=\int^{\infty}_{\infty} \frac{d\Omega}{2\pi}  \Big( \vcenter{\hbox{\includegraphics[width=0.25\columnwidth,keepaspectratio]{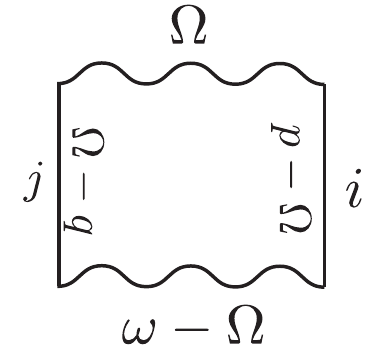}}} \Big) \\ \nonumber
&= \int^{\infty}_{-\infty} \frac{d\Omega}{2\pi} \frac{(g \cos(\theta))^4} {N^2}G^R_{\tilde\phi} (\Omega) G^R_{\tilde\phi} (\omega-\Omega) \\ \nonumber
&\times (G^W_{\tilde{f}}(p-\Omega)+G^W_{\tilde{f}^\dagger}(p-\Omega)) (G^W_{\tilde{f}}(\Omega-q)+G^W_{\tilde{f}^\dagger}(\Omega-q)).
\end{align}
In this diagram an implicit sum over all four possible orientations of Fermionic arrows is assumed. To this order $R^{i,j}$ is independent of $i,j$. A longer and more detailed expression for Eq.\eqref{m1} (using Eq.\eqref{rmj}) is given in the Appendix.\ref{a1}.

The right hand side of Eq.\eqref{m1} (also see Appendix.\ref{a1}) is proportional to $1/N$. This shows that the thermal state is not chaotic in the $N\rightarrow \infty$ limit\footnote{ This does not hold for single highly excited states (as opposed to the thermal state) with a large fixed total spin. See Refs.\onlinecite{rey,1807a}.}. 

The leading order rung diagram shown above does \textit{not} involve the interaction vertex unique to the super radiant phase (Fig.\ref{nvtx}) (though they are present at higher orders and will be discussed in the results section). This suggests that $\mathcal{C}_{\tilde\sigma_z}$ might be blind to some features of the super-radiant phase. This is a special and fine-tuned feature of $\tilde{\sigma}_z$. In contrast, \textit{leading} order expressions for OTOCs associated with other spin operators (e.g. $\sigma_z$) involve also diagrams that are nonzero only in the super radiant phase.

The two Wightman functions (last two terms) in Eq.\eqref{rmj} make $R^{i,j}\propto \frac{1}{\cosh^2(\beta\tilde\omega_z/2)}$. This already implies that the spin scrambling rate is strongly suppressed at very low temperatures $\beta\tilde\omega_z\gg 1$.

\subsection{Diagrammatic form of $\mathcal{C}_{\tilde\phi}(t)$}\label{cphi}
Similar to $f_{\tilde\sigma_z}(\omega,p)$, $f_{\tilde\phi}(\omega,p)$ can be defined as,
\begin{align}
\mathcal{C}_{\tilde\phi}(\omega)=\int_{-\infty}^\infty \frac{dp}{2\pi} f_{\tilde\phi}(\omega,p).
\end{align}
A integral equation for $f_{\tilde\phi}(\omega,p)$ is shown in Fig.\ref{b-s}. This equation can be explicitly written as,
\begin{align}
&f_{\tilde\phi}(\omega,p)= - G^R_{\tilde\phi} (p) G^R_{\tilde\phi} (\omega-p) \nonumber \\
&\times\Big[1+ \int_{-\infty}^{\infty} \frac{dq}{2\pi}  R_{\tilde\phi}(\omega,p,q)  f_{\tilde\phi}(\omega,p)\Big].
\end{align}
As in the previous case, we drop the first term to get,
\begin{align}\label{m2}
&f_{\tilde\phi}(\omega,p)= - G^R_{\tilde\phi} (p) G^R_{\tilde\phi} (\omega-p) \nonumber \\
&\times \int_{-\infty}^{\infty} \frac{dq}{2\pi}  R_{\tilde\phi}(\omega,p,q)  f_{\tilde\phi}(\omega,p).
\end{align}

 A total of $24$ diagrams now contribute to the leading order approximation of $R_{\tilde\phi}(\omega,p,q)$ (shown in Fig.\ref{Boson_rung.pdf}). In Fig.\ref{Boson_rung.pdf} we have dropped all diagrams with identical Wightman functions, this is because parallel and crossed leg versions of such diagrams cancel out each other (contribute with the same magnitude and opposite sign), for example;
\begin{align}
\vcenter{\hbox{\includegraphics[width=0.4\columnwidth,keepaspectratio]{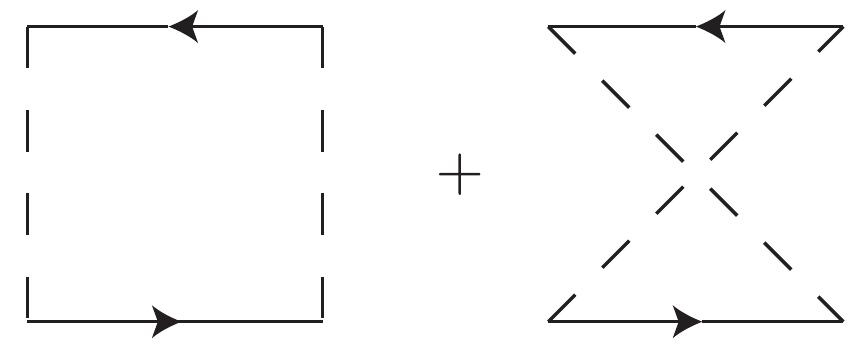}}}=0.
\end{align}
It is interesting to note that all $\tilde{\phi}$ rung functions $R_{\tilde\phi}(\omega,p,q)$ shown in Fig.\ref{Boson_rung.pdf} are also present as sub-leading corrections to $R_{\tilde{\sigma_z}}(\omega,p,q)$.

Similar to the previous case, temperature scaling of the first two diagrams in Fig.\ref{Boson_rung.pdf} is $\frac{1}{\cosh^2(\beta\tilde\omega_z/2)}$. However that last four diagrams which are only nonzero in the super-radiant phase scale with $\frac{1}{\cosh(\beta\tilde\omega_z/2)}$. Both of these terms still decay exponentially as $\beta\tilde\omega_z\rightarrow \infty$. Nonetheless there exists an intermediate temperature regime, where the last four diagrams dominate.

A detailed expression for Eq.\eqref{m2}, using the diagrams in Fig.\ref{Boson_rung.pdf} is given in the Appendix.\ref{a1}.

We again note that the right hand side of Eq.\eqref{m2} is proportional to $1/N$ (scrambling is a finite $N$ effect).

  \begin{figure}[t]
  \vspace{0.05in}
  \includegraphics[width=0.94\columnwidth]{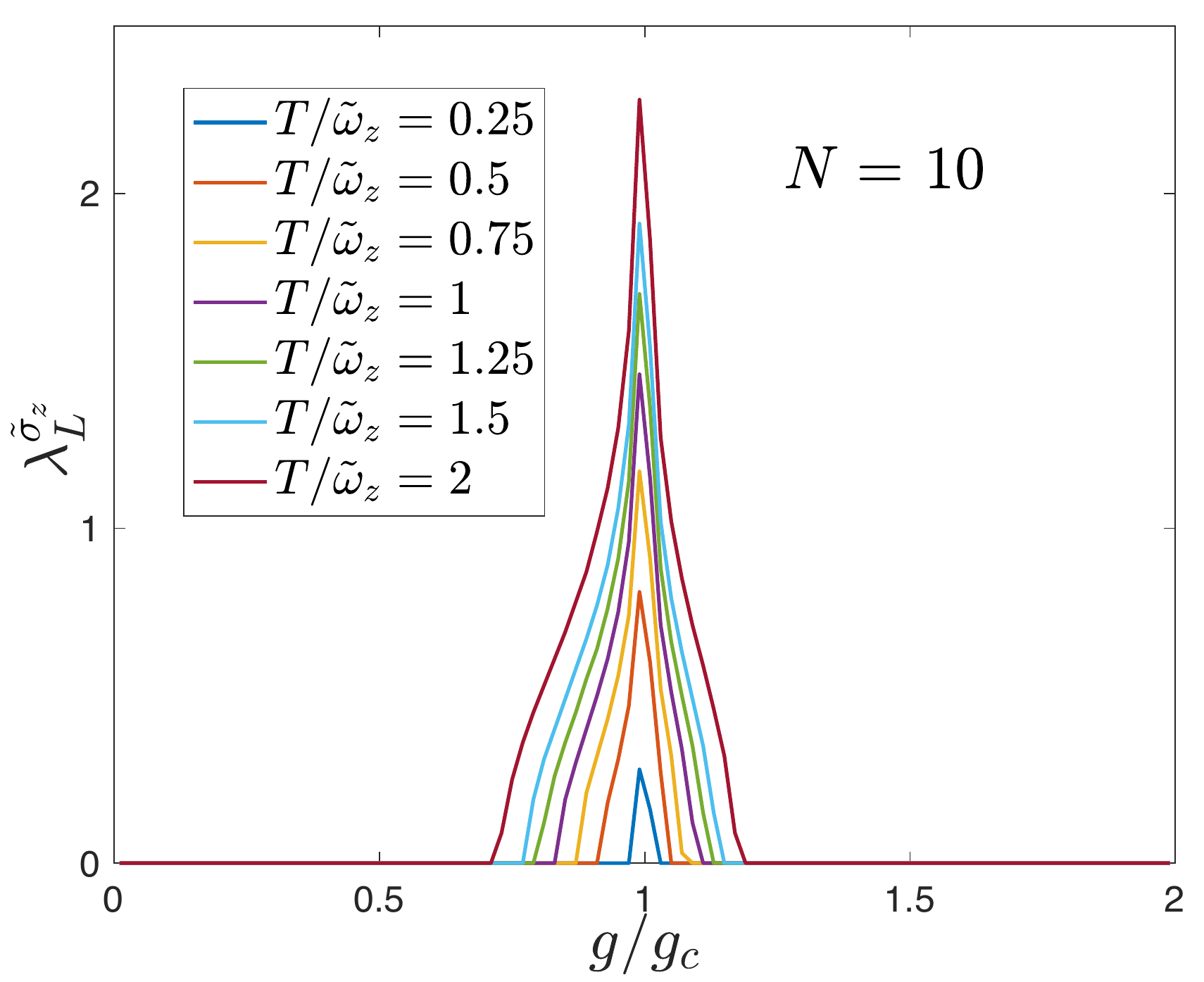}
  \caption{ Scrambling rate of $\tilde{\sigma}_z$ a as function of $g/g_c$, for multiple fixed values of $T/\tilde\omega_z$. Note that $g_c$ is temperature dependent (Eq.\eqref{gc}). Here $\omega_0=\omega_z=1$.}
  \label{Fig1.pdf}
\end{figure}
\section{Results and Discussion}\label{reslt}
 \begin{figure}[ht]
  \includegraphics[width=\columnwidth]{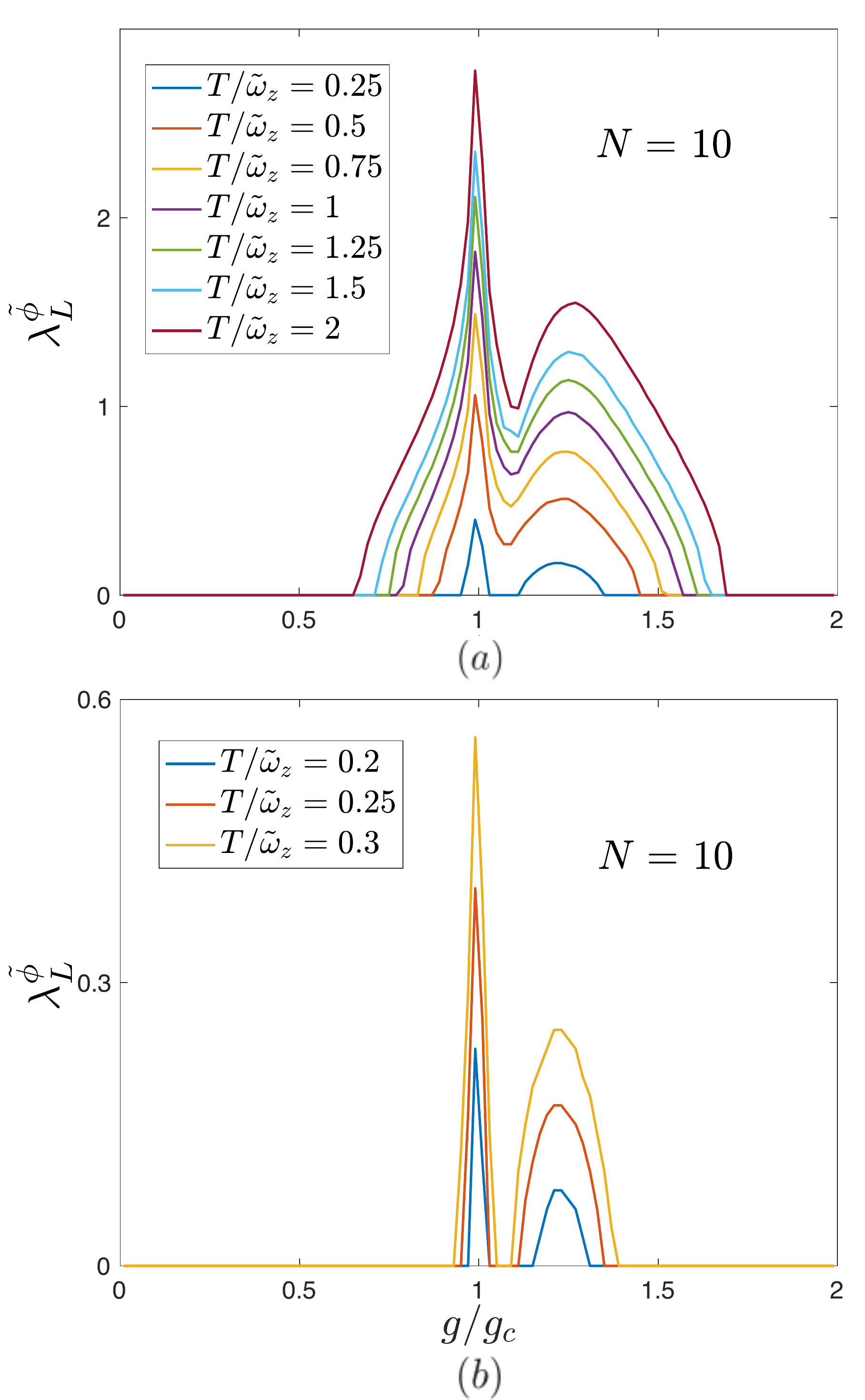}
  \caption{Boson scrambling rate as a function of $g/g_c$, for multiple fixed values of $T/\tilde\omega_z$; (a) over a broad range of temperature and (b) at low temperatures. Note that $g_c$ is temperature dependent (Eq.\eqref{gc}). Here $\omega_0=\omega_z=1$.}
  \label{Fig2.pdf}
\end{figure}

In this section we use the integral Eqs.\eqref{m1} and \eqref{m2} to compute $\lambda_L$. To solve these equations numerically, we discretize them as matrix equations of the following form,
\begin{align}\label{mxxe}
\sum_q M_{p,q}(\omega) f_q(\omega)=0.
\end{align}
In fact since the leading order expressions for Wightman functions (Eq.\eqref{wmf}) involves delta functions, the integral equations are straightforward to discretize (see Appendix.\ref{a1}).

 A nonzero solution of Eq.\eqref{mxxe} along the positive imaginary axis, $\omega=i \lambda$, indicates an exponential growth of the corresponding OTOC\cite{otoc1}. The scrambling rate $\lambda_L$ is then given by the largest $\lambda$ where such a solution exists.

Details of the method used to find $\lambda_L$ is given in the Appendix.\ref{a2} . For simplicity, in all our numerical results, we set $\omega_0=\omega_z=1$.

 The $\tilde{\sigma}_z$ scrambling rate $\lambda_L^{\tilde{\sigma}_z}$ as a function of the coupling strength $g$, at multiple fixed values of $T/\tilde\omega_z$ is plotted in Fig.\ref{Fig1.pdf}. As shown in the figure, at low temperatures $T\ll \tilde{\omega}_z$, chaotic behavior is limited to the close vicinity of the critical point $g\approx g_c$. As the temperature is increased the magnitude of  $\lambda_L^{\tilde{\sigma}_z}$ as well as the size of the region over which $\lambda_L^{\tilde{\sigma}_z}\neq 0$, are both monotonically increased. $\lambda_L^{\tilde{\sigma}_z}$ is nonzero in \textit{both} the normal and super-radiant phases.

 Similarly, the Bosonic scrambling rate $\lambda_L^{\tilde{\phi}}$  is plotted in Fig.\ref{Fig2.pdf}. As shown in Fig.\ref{Fig2.pdf}(b) and in contrast to the previous case ($\lambda_L^{\tilde{\sigma}_z}$), at low temperature $T\ll \tilde{\omega}_z$ chaotic behavior is not limited to the vicinity of the critical point, instead it now also includes a finite region deep within the super-radiant phase. Similar to the previous case, as the temperature is increased the magnitude of $\lambda_L^{\tilde\phi}$ as well as the size of the region over which $\lambda_L^{\tilde\phi} \neq 0$, are both increased. However, note that in this case, chaotic behavior is manifestly stronger in the super-radiant phase. In particular size of the chaotic region is significantly larger in super-radiant phase.

As shown in Figs.\ref{Fig1.pdf} and \ref{Fig2.pdf}, $\lambda_L^{\tilde{\phi}}$ and $\lambda_L^{\tilde{\sigma}_z}$ are similar to each other in the normal-phase, whereas they look qualitatively different in the super-radiance phase. Their difference in the super-radiant phase can be attributed to the fact that the leading order diagrams used to compute $\lambda_L^{\tilde{\sigma}_z}$ do \textit{not} involve the interaction vertex unique to the super-radiant phase (see Section\ref{csz}). For this reason, signatures of chaos unique to the super-radiant phase are not manifest in $\lambda_L^{\tilde{\sigma}_z}$. In contrast, diagrams used to compute $\lambda_L^{\tilde{\phi}}$, explicitly involve diagrams special to the super-radiant phase (the last four diagrams in Fig.\ref{Boson_rung.pdf}) and hence, $\lambda_L^{\tilde{\phi}}$ is sensitive to distinctive properties of the super-radiant phase. In fact, the ``dome" like feature displayed in Fig.\ref{Fig2.pdf} is directly associated with the last four diagrams of Fig.\ref{Boson_rung.pdf}. To check this, we have artificially set the value of these diagrams to zero and confirmed that the resulting behavior is almost \textit{identical} to Fig.\ref{Fig1.pdf}. Therefore, we believe that $\lambda_L^{\tilde{\phi}}$ (as opposed to $\lambda_L^{\tilde{\sigma}_z}$) describes the generic chaotic features of the DM and that the behavior of $\tilde{\sigma}_z$ is fine-tuned (as discussed in Section\ref{csz}) and does not represent the generic chaotic behavior of this system. However, their comparison provides a useful tool to identity chaotic features unique to the super-radiant phase.

Note that since $\tilde{\phi}$ and $\tilde{\sigma}_z$ are coupled, all diagrams giving rise to exponential behavior for one operator (say $\tilde{\phi}$) also appear as part of the diagrams for the other operator ($\tilde{\sigma}_z$). Therefore, one might be led to conclude that the two scrambling rates have to be equal. However, note that these diagrams can be of different orders in perturbation theory. In fact as mentioned in section \ref{cphi} all diagrams involved in calculating  $\lambda_L^{\tilde{\phi}}$ are also present as \textit{sub-leading} ($1/N^2$) corrections to $\lambda_L^{\tilde{\sigma}_z}$. This suggests an interesting situation where,
 \begin{align}
 \mathcal{C}_{\tilde{\sigma_z}}(t)\sim \frac{c_1}{N} e^{\lambda_L^{\tilde{\sigma}_z} t}+ \frac{c_2}{N^2} e^{\lambda_L^{\tilde{\phi}} t} + . . .\quad .
 \end{align}
 So for small values of $N$, either exponent ($\lambda_L^{\tilde{\phi}}$ or $\lambda_L^{\tilde{\sigma}_z}$) could dominate the early time behavior. However, for large enough $N$ the early time behavior is determined by the first term. Therefore, despite the fact that $\tilde{\phi}$ and $\tilde{\sigma}_z$ are coupled, the scrambling rates associated with them are different.

 \begin{figure}[t]
  \includegraphics[width=\columnwidth]{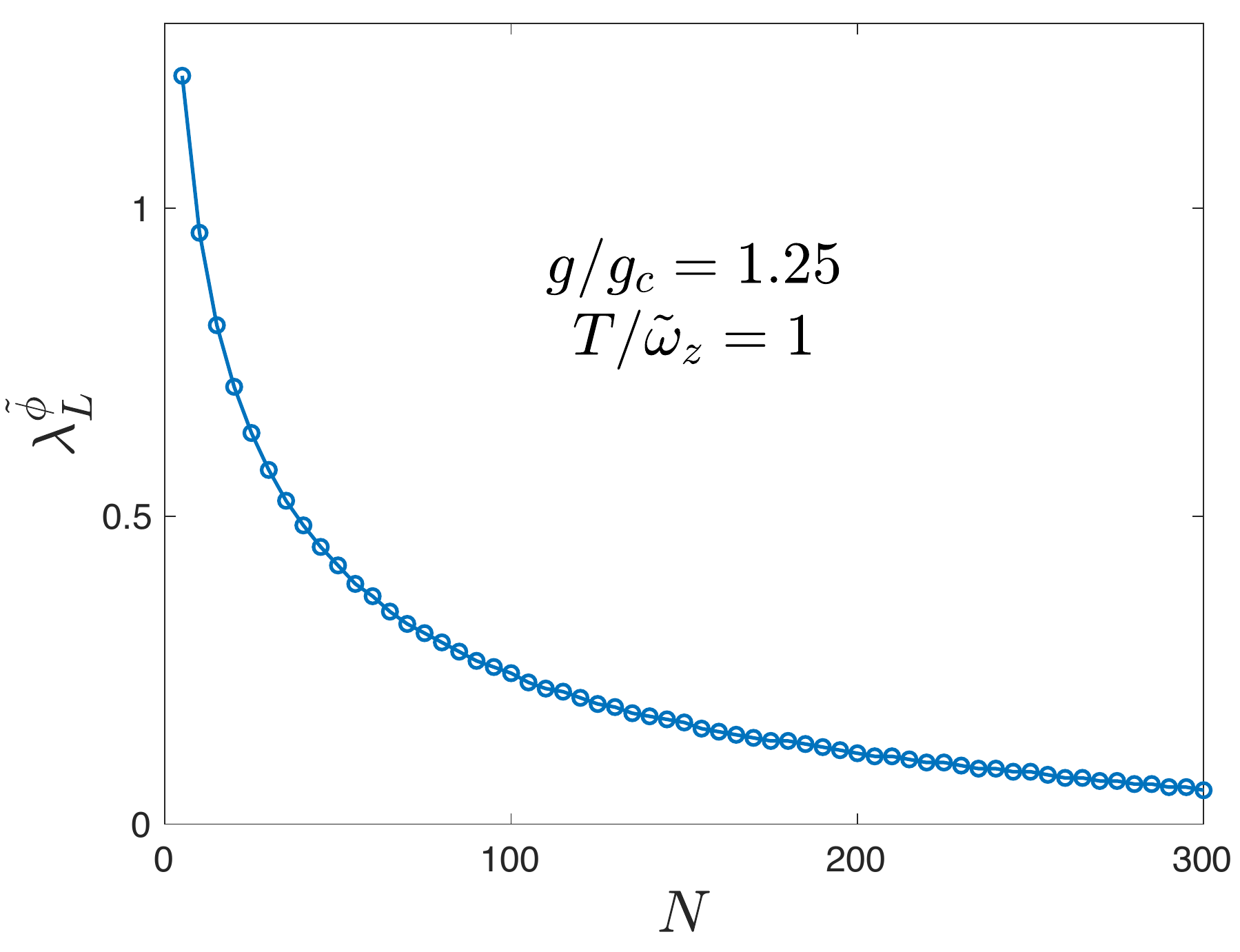}
  \caption{Boson scrambling rate as a function of $N$, at fixed values of $g/g_c$ and $T/\tilde{\omega}_z$. Note that $g_c$ is temperature dependent (Eq.\eqref{gc}). Here $\omega_0=\omega_z=1$.}
  \label{Ndep}
\end{figure}

Bosonic scrambling rate $\lambda_L^{\tilde{\phi}}$ as a function of $N$ at fixed valuex of $g/g_c$ and $T/\tilde{\omega}_z$ is plotted in Fig\ref{Ndep}. As expected the $\lambda_L^{\tilde{\phi}}$ is a monotonically decreasing function of $N$. At large values of $N$, $\lambda_L^{\tilde{\phi}}$ becomes zero. This is expected since in the  $N\rightarrow\infty$ limit the system becomes integrable.

Note that our results clearly indicate that $\lambda_L$ can be nonzero in the normal phase. This might seem to be counter intuitive according to the conventional wisdom\cite{dc1,dc2} based zero temperature studies of the DM. However, note that our critical value of coupling $g_c (T)$ (Eq.\eqref{gc}) is temperature dependent and for this reason regions in the phase diagram where $g_c(0)<g<g_c(T)$ are considered as normal phase in our paper.  Moreover, multiple more recent semiclassical studies of chaos in the DM have all found that chaos also exists in the normal phase, specially at high energies\cite{dc3,dc4,dc5,dc6,dc8} (in our case this translates into high temperatures).

Another potentially confusing point is that Figs.\ref{Fig1.pdf} and \ref{Fig2.pdf} show that $\lambda_L$ becomes zero above some value of $g/g_c$ in the super-radiant phase. To understand this note that, at large values of $g\gg g_c$ the system approaches integrability again. This issue has already been addressed in Refs.\onlinecite{dc1,dc2}. There, it is shown that in the super-radiant phase as one increases $g/g_c$, the lower part of the spectrum becomes regular. The size of the regular part of the spectrum increases with $g/g_c$. In our results this shows as $\lambda$ being zero at low temperatures and being nonzero at high temperatures (see e.g. Fig.\ref{Fig2.pdf}).

 \section{Summary and Conclusion}\label{d&c}

 We used the Majorana representation of spin $1/2$ to obtain an effective theory for the DM model in the super-radiant phase (Eqs.\eqref{thetae} and \eqref{ef2}). We found a new set of natural variables ($\tilde{\sigma}_z$ and $\tilde\phi$) and a new interaction vertex (Fig.\ref{nvtx}) distinguishing normal and super-radiant phases. This effective theory was then used to compute the scrambling rate $\lambda_L$ associated with $\tilde{\sigma}_z$ and $\tilde\phi$. At low temperatures the chaotic behavior is limited to (a) a region within the super-radiant phase and (b) vicinity of the critical point. At high temperatures $\lambda_L$ becomes nonzero in an extended region that includes both the normal and super-radiant phases (see Figs.\ref{Fig1.pdf} and \ref{Fig2.pdf}). We identified the dome like feature of $\lambda_L^{\tilde{\phi}}$  (shown in Fig.\ref{Fig2.pdf}) as the key feature distinguishing chaotic behavior in normal and super-radiant phases. We discussed and compared our results with the existing semiclassical studies of chaos in the DM.

 Experimental attempts to measure $\lambda_L$ in the DM are already underway\cite{exp0,exp1}. This can potentially make our results to be of short-term experimental relevance. Finally, we note that our formalism can be easily extended to various generalizations of the DM\cite{dmrev}. Several interesting candidates already exist in the literature\cite{ext1,ext2,ext3,ext4}.

{\it Note added.} Upon finishing the manuscript we become aware of two recent preprints ~\cite{1807a,rey} that has performed a numerical evaluation of Bosonic OTOC for some specific eigenstates of the DM.

 \section*{Acknowledgement}
 We are grateful to Jay Sau, Brian Swingle, Moahmmad Hafezi and Victor Galitski for enlightening discussions and valuable comments on the manuscript. Y.A. was supported by JQI-NSF-PFC and  the National Science Foundation NSF DMR-1555135. A.L. was supported by JQI-PFC-UMD.

\bibliography{library}

\begin{thebibliography}{64}%
\makeatletter
\providecommand \@ifxundefined [1]{%
 \@ifx{#1\undefined}
}%
\providecommand \@ifnum [1]{%
 \ifnum #1\expandafter \@firstoftwo
 \else \expandafter \@secondoftwo
 \fi
}%
\providecommand \@ifx [1]{%
 \ifx #1\expandafter \@firstoftwo
 \else \expandafter \@secondoftwo
 \fi
}%
\providecommand \natexlab [1]{#1}%
\providecommand \enquote  [1]{``#1''}%
\providecommand \bibnamefont  [1]{#1}%
\providecommand \bibfnamefont [1]{#1}%
\providecommand \citenamefont [1]{#1}%
\providecommand \href@noop [0]{\@secondoftwo}%
\providecommand \href [0]{\begingroup \@sanitize@url \@href}%
\providecommand \@href[1]{\@@startlink{#1}\@@href}%
\providecommand \@@href[1]{\endgroup#1\@@endlink}%
\providecommand \@sanitize@url [0]{\catcode `\\12\catcode `\$12\catcode
  `\&12\catcode `\#12\catcode `\^12\catcode `\_12\catcode `\%12\relax}%
\providecommand \@@startlink[1]{}%
\providecommand \@@endlink[0]{}%
\providecommand \url  [0]{\begingroup\@sanitize@url \@url }%
\providecommand \@url [1]{\endgroup\@href {#1}{\urlprefix }}%
\providecommand \urlprefix  [0]{URL }%
\providecommand \Eprint [0]{\href }%
\providecommand \doibase [0]{http://dx.doi.org/}%
\providecommand \selectlanguage [0]{\@gobble}%
\providecommand \bibinfo  [0]{\@secondoftwo}%
\providecommand \bibfield  [0]{\@secondoftwo}%
\providecommand \translation [1]{[#1]}%
\providecommand \BibitemOpen [0]{}%
\providecommand \bibitemStop [0]{}%
\providecommand \bibitemNoStop [0]{.\EOS\space}%
\providecommand \EOS [0]{\spacefactor3000\relax}%
\providecommand \BibitemShut  [1]{\csname bibitem#1\endcsname}%
\let\auto@bib@innerbib\@empty
\bibitem [{\citenamefont {Larkin}\ and\ \citenamefont
  {Ovchinnikov}(1969)}]{larkin}%
  \BibitemOpen
  \bibfield  {author} {\bibinfo {author} {\bibfnamefont {A.}~\bibnamefont
  {Larkin}}\ and\ \bibinfo {author} {\bibfnamefont {Y.~N.}\ \bibnamefont
  {Ovchinnikov}},\ }\href@noop {} {\bibfield  {journal} {\bibinfo  {journal}
  {Sov Phys JETP}\ }\textbf {\bibinfo {volume} {28}},\ \bibinfo {pages} {1200}
  (\bibinfo {year} {1969})}\BibitemShut {NoStop}%
\bibitem [{\citenamefont {Kitaev}(2015)}]{Kitaev}%
  \BibitemOpen
  \bibfield  {author} {\bibinfo {author} {\bibfnamefont {A.}~\bibnamefont
  {Kitaev}},\ }\href@noop {} {\enquote {\bibinfo {title} {A simple model of
  quantum holography talk at kitp},}\ } (\bibinfo {year} {2015})\BibitemShut
  {NoStop}%
\bibitem [{\citenamefont {Maldacena}\ \emph {et~al.}(2016)\citenamefont
  {Maldacena}, \citenamefont {Shenker},\ and\ \citenamefont
  {Stanford}}]{Maldacena}%
  \BibitemOpen
  \bibfield  {author} {\bibinfo {author} {\bibfnamefont {J.}~\bibnamefont
  {Maldacena}}, \bibinfo {author} {\bibfnamefont {S.~H.}\ \bibnamefont
  {Shenker}}, \ and\ \bibinfo {author} {\bibfnamefont {D.}~\bibnamefont
  {Stanford}},\ }\href {\doibase 10.1007/JHEP08(2016)106} {\bibfield  {journal}
  {\bibinfo  {journal} {Journal of High Energy Physics}\ }\textbf {\bibinfo
  {volume} {2016}},\ \bibinfo {pages} {106} (\bibinfo {year}
  {2016})}\BibitemShut {NoStop}%
\bibitem [{Note1()}]{Note1}%
  \BibitemOpen
  \bibinfo {note} {Though the relation between $\lambda _L$ and the classical
  Lyapunov exponent is subtle and not straightforward (see Ref.\protect
  \rev@citealpnum {Efim1}).}\BibitemShut {Stop}%
\bibitem [{\citenamefont {Patel}\ and\ \citenamefont {Sachdev}(2017)}]{otoc1}%
  \BibitemOpen
  \bibfield  {author} {\bibinfo {author} {\bibfnamefont {A.~A.}\ \bibnamefont
  {Patel}}\ and\ \bibinfo {author} {\bibfnamefont {S.}~\bibnamefont
  {Sachdev}},\ }\href {\doibase 10.1073/pnas.1618185114} {\bibfield  {journal}
  {\bibinfo  {journal} {Proceedings of the National Academy of Sciences}\
  }\textbf {\bibinfo {volume} {114}},\ \bibinfo {pages} {1844} (\bibinfo {year}
  {2017})},\ \Eprint
  {http://arxiv.org/abs/http://www.pnas.org/content/114/8/1844.full.pdf}
  {http://www.pnas.org/content/114/8/1844.full.pdf} \BibitemShut {NoStop}%
\bibitem [{\citenamefont {Chowdhury}\ and\ \citenamefont
  {Swingle}(2017)}]{otoc2}%
  \BibitemOpen
  \bibfield  {author} {\bibinfo {author} {\bibfnamefont {D.}~\bibnamefont
  {Chowdhury}}\ and\ \bibinfo {author} {\bibfnamefont {B.}~\bibnamefont
  {Swingle}},\ }\href {\doibase 10.1103/PhysRevD.96.065005} {\bibfield
  {journal} {\bibinfo  {journal} {Phys. Rev. D}\ }\textbf {\bibinfo {volume}
  {96}},\ \bibinfo {pages} {065005} (\bibinfo {year} {2017})}\BibitemShut
  {NoStop}%
\bibitem [{\citenamefont {Banerjee}\ and\ \citenamefont
  {Altman}(2017)}]{otoc3}%
  \BibitemOpen
  \bibfield  {author} {\bibinfo {author} {\bibfnamefont {S.}~\bibnamefont
  {Banerjee}}\ and\ \bibinfo {author} {\bibfnamefont {E.}~\bibnamefont
  {Altman}},\ }\href {\doibase 10.1103/PhysRevB.95.134302} {\bibfield
  {journal} {\bibinfo  {journal} {Phys. Rev. B}\ }\textbf {\bibinfo {volume}
  {95}},\ \bibinfo {pages} {134302} (\bibinfo {year} {2017})}\BibitemShut
  {NoStop}%
\bibitem [{\citenamefont {Gu}\ \emph {et~al.}(2017)\citenamefont {Gu},
  \citenamefont {Lucas},\ and\ \citenamefont {Qi}}]{otoc4}%
  \BibitemOpen
  \bibfield  {author} {\bibinfo {author} {\bibfnamefont {Y.}~\bibnamefont
  {Gu}}, \bibinfo {author} {\bibfnamefont {A.}~\bibnamefont {Lucas}}, \ and\
  \bibinfo {author} {\bibfnamefont {X.-L.}\ \bibnamefont {Qi}},\ }\href
  {\doibase 10.21468/SciPostPhys.2.3.018} {\bibfield  {journal} {\bibinfo
  {journal} {SciPost Phys.}\ }\textbf {\bibinfo {volume} {2}},\ \bibinfo
  {pages} {018} (\bibinfo {year} {2017})}\BibitemShut {NoStop}%
\bibitem [{\citenamefont {Bohrdt}\ \emph {et~al.}(2017)\citenamefont {Bohrdt},
  \citenamefont {Mendl}, \citenamefont {Endres},\ and\ \citenamefont
  {Knap}}]{otoc5}%
  \BibitemOpen
  \bibfield  {author} {\bibinfo {author} {\bibfnamefont {A.}~\bibnamefont
  {Bohrdt}}, \bibinfo {author} {\bibfnamefont {C.}~\bibnamefont {Mendl}},
  \bibinfo {author} {\bibfnamefont {M.}~\bibnamefont {Endres}}, \ and\ \bibinfo
  {author} {\bibfnamefont {M.}~\bibnamefont {Knap}},\ }\href@noop {} {\bibfield
   {journal} {\bibinfo  {journal} {New Journal of Physics}\ }\textbf {\bibinfo
  {volume} {19}},\ \bibinfo {pages} {063001} (\bibinfo {year}
  {2017})}\BibitemShut {NoStop}%
\bibitem [{\citenamefont {Patel}\ \emph {et~al.}(2017)\citenamefont {Patel},
  \citenamefont {Chowdhury}, \citenamefont {Sachdev},\ and\ \citenamefont
  {Swingle}}]{otoc6}%
  \BibitemOpen
  \bibfield  {author} {\bibinfo {author} {\bibfnamefont {A.~A.}\ \bibnamefont
  {Patel}}, \bibinfo {author} {\bibfnamefont {D.}~\bibnamefont {Chowdhury}},
  \bibinfo {author} {\bibfnamefont {S.}~\bibnamefont {Sachdev}}, \ and\
  \bibinfo {author} {\bibfnamefont {B.}~\bibnamefont {Swingle}},\ }\href
  {\doibase 10.1103/PhysRevX.7.031047} {\bibfield  {journal} {\bibinfo
  {journal} {Phys. Rev. X}\ }\textbf {\bibinfo {volume} {7}},\ \bibinfo {pages}
  {031047} (\bibinfo {year} {2017})}\BibitemShut {NoStop}%
\bibitem [{\citenamefont {Blake}\ \emph {et~al.}(2017)\citenamefont {Blake},
  \citenamefont {Davison},\ and\ \citenamefont {Sachdev}}]{otoc7}%
  \BibitemOpen
  \bibfield  {author} {\bibinfo {author} {\bibfnamefont {M.}~\bibnamefont
  {Blake}}, \bibinfo {author} {\bibfnamefont {R.~A.}\ \bibnamefont {Davison}},
  \ and\ \bibinfo {author} {\bibfnamefont {S.}~\bibnamefont {Sachdev}},\ }\href
  {\doibase 10.1103/PhysRevD.96.106008} {\bibfield  {journal} {\bibinfo
  {journal} {Phys. Rev. D}\ }\textbf {\bibinfo {volume} {96}},\ \bibinfo
  {pages} {106008} (\bibinfo {year} {2017})}\BibitemShut {NoStop}%
\bibitem [{\citenamefont {Kukuljan}\ \emph {et~al.}(2017)\citenamefont
  {Kukuljan}, \citenamefont {Grozdanov},\ and\ \citenamefont {Prosen}}]{otoc8}%
  \BibitemOpen
  \bibfield  {author} {\bibinfo {author} {\bibfnamefont {I.}~\bibnamefont
  {Kukuljan}}, \bibinfo {author} {\bibfnamefont {S.~c.~v.}\ \bibnamefont
  {Grozdanov}}, \ and\ \bibinfo {author} {\bibfnamefont {T.~c.~v.}\
  \bibnamefont {Prosen}},\ }\href {\doibase 10.1103/PhysRevB.96.060301}
  {\bibfield  {journal} {\bibinfo  {journal} {Phys. Rev. B}\ }\textbf {\bibinfo
  {volume} {96}},\ \bibinfo {pages} {060301} (\bibinfo {year}
  {2017})}\BibitemShut {NoStop}%
\bibitem [{\citenamefont {Nahum}\ \emph {et~al.}(2018)\citenamefont {Nahum},
  \citenamefont {Vijay},\ and\ \citenamefont {Haah}}]{otoc9}%
  \BibitemOpen
  \bibfield  {author} {\bibinfo {author} {\bibfnamefont {A.}~\bibnamefont
  {Nahum}}, \bibinfo {author} {\bibfnamefont {S.}~\bibnamefont {Vijay}}, \ and\
  \bibinfo {author} {\bibfnamefont {J.}~\bibnamefont {Haah}},\ }\href {\doibase
  10.1103/PhysRevX.8.021014} {\bibfield  {journal} {\bibinfo  {journal} {Phys.
  Rev. X}\ }\textbf {\bibinfo {volume} {8}},\ \bibinfo {pages} {021014}
  (\bibinfo {year} {2018})}\BibitemShut {NoStop}%
\bibitem [{\citenamefont {Lin}\ and\ \citenamefont {Motrunich}(2018)}]{otoc10}%
  \BibitemOpen
  \bibfield  {author} {\bibinfo {author} {\bibfnamefont {C.-J.}\ \bibnamefont
  {Lin}}\ and\ \bibinfo {author} {\bibfnamefont {O.~I.}\ \bibnamefont
  {Motrunich}},\ }\href {\doibase 10.1103/PhysRevB.97.144304} {\bibfield
  {journal} {\bibinfo  {journal} {Phys. Rev. B}\ }\textbf {\bibinfo {volume}
  {97}},\ \bibinfo {pages} {144304} (\bibinfo {year} {2018})}\BibitemShut
  {NoStop}%
\bibitem [{\citenamefont {Syzranov}\ \emph {et~al.}(2018)\citenamefont
  {Syzranov}, \citenamefont {Gorshkov},\ and\ \citenamefont
  {Galitski}}]{otoc11}%
  \BibitemOpen
  \bibfield  {author} {\bibinfo {author} {\bibfnamefont {S.~V.}\ \bibnamefont
  {Syzranov}}, \bibinfo {author} {\bibfnamefont {A.~V.}\ \bibnamefont
  {Gorshkov}}, \ and\ \bibinfo {author} {\bibfnamefont {V.}~\bibnamefont
  {Galitski}},\ }\href {\doibase 10.1103/PhysRevB.97.161114} {\bibfield
  {journal} {\bibinfo  {journal} {Phys. Rev. B}\ }\textbf {\bibinfo {volume}
  {97}},\ \bibinfo {pages} {161114} (\bibinfo {year} {2018})}\BibitemShut
  {NoStop}%
\bibitem [{\citenamefont {Plamadeala}\ and\ \citenamefont
  {Fradkin}(2018)}]{otoc12}%
  \BibitemOpen
  \bibfield  {author} {\bibinfo {author} {\bibfnamefont {E.}~\bibnamefont
  {Plamadeala}}\ and\ \bibinfo {author} {\bibfnamefont {E.}~\bibnamefont
  {Fradkin}},\ }\href@noop {} {\enquote {\bibinfo {title} {Scrambling in the
  quantum lifshitz model},}\ } (\bibinfo {year} {2018}),\ \Eprint
  {http://arxiv.org/abs/arXiv:1802.07268} {arXiv:1802.07268} \BibitemShut
  {NoStop}%
\bibitem [{\citenamefont {Klug}\ \emph {et~al.}(2017)\citenamefont {Klug},
  \citenamefont {Scheurer},\ and\ \citenamefont {Schmalian}}]{otoc13}%
  \BibitemOpen
  \bibfield  {author} {\bibinfo {author} {\bibfnamefont {M.~J.}\ \bibnamefont
  {Klug}}, \bibinfo {author} {\bibfnamefont {M.~S.}\ \bibnamefont {Scheurer}},
  \ and\ \bibinfo {author} {\bibfnamefont {J.}~\bibnamefont {Schmalian}},\
  }\href@noop {} {\enquote {\bibinfo {title} {Hierarchy of information
  scrambling, thermalization, and hydrodynamic flow in graphene},}\ } (\bibinfo
  {year} {2017}),\ \Eprint {http://arxiv.org/abs/arXiv:1712.08813}
  {arXiv:1712.08813} \BibitemShut {NoStop}%
\bibitem [{\citenamefont {Bentsen}\ \emph {et~al.}(2018)\citenamefont
  {Bentsen}, \citenamefont {Gu},\ and\ \citenamefont {Lucas}}]{otoc14}%
  \BibitemOpen
  \bibfield  {author} {\bibinfo {author} {\bibfnamefont {G.}~\bibnamefont
  {Bentsen}}, \bibinfo {author} {\bibfnamefont {Y.}~\bibnamefont {Gu}}, \ and\
  \bibinfo {author} {\bibfnamefont {A.}~\bibnamefont {Lucas}},\ }\href@noop {}
  {\enquote {\bibinfo {title} {Fast scrambling on sparse graphs},}\ } (\bibinfo
  {year} {2018}),\ \Eprint {http://arxiv.org/abs/arXiv:1805.08215}
  {arXiv:1805.08215} \BibitemShut {NoStop}%
\bibitem [{\citenamefont {Vijay}\ and\ \citenamefont
  {Vishwanath}(2018)}]{otoc15}%
  \BibitemOpen
  \bibfield  {author} {\bibinfo {author} {\bibfnamefont {S.}~\bibnamefont
  {Vijay}}\ and\ \bibinfo {author} {\bibfnamefont {A.}~\bibnamefont
  {Vishwanath}},\ }\href@noop {} {\enquote {\bibinfo {title}
  {Finite-temperature scrambling of a random hamiltonian},}\ } (\bibinfo {year}
  {2018}),\ \Eprint {http://arxiv.org/abs/arXiv:1803.08483} {arXiv:1803.08483}
  \BibitemShut {NoStop}%
\bibitem [{\citenamefont {Werman}\ \emph {et~al.}(2017)\citenamefont {Werman},
  \citenamefont {Kivelson},\ and\ \citenamefont {Berg}}]{otoc16}%
  \BibitemOpen
  \bibfield  {author} {\bibinfo {author} {\bibfnamefont {Y.}~\bibnamefont
  {Werman}}, \bibinfo {author} {\bibfnamefont {S.~A.}\ \bibnamefont
  {Kivelson}}, \ and\ \bibinfo {author} {\bibfnamefont {E.}~\bibnamefont
  {Berg}},\ }\href@noop {} {\enquote {\bibinfo {title} {Quantum chaos in an
  electron-phonon bad metal},}\ } (\bibinfo {year} {2017}),\ \Eprint
  {http://arxiv.org/abs/arXiv:1705.07895} {arXiv:1705.07895} \BibitemShut
  {NoStop}%
\bibitem [{\citenamefont {D\'ora}\ \emph {et~al.}(2017)\citenamefont {D\'ora},
  \citenamefont {Werner},\ and\ \citenamefont {Moca}}]{otoc17}%
  \BibitemOpen
  \bibfield  {author} {\bibinfo {author} {\bibfnamefont {B.}~\bibnamefont
  {D\'ora}}, \bibinfo {author} {\bibfnamefont {M.~A.}\ \bibnamefont {Werner}},
  \ and\ \bibinfo {author} {\bibfnamefont {C.~u. u. u. u. P. m.~c.}\
  \bibnamefont {Moca}},\ }\href {\doibase 10.1103/PhysRevB.96.155116}
  {\bibfield  {journal} {\bibinfo  {journal} {Phys. Rev. B}\ }\textbf {\bibinfo
  {volume} {96}},\ \bibinfo {pages} {155116} (\bibinfo {year}
  {2017})}\BibitemShut {NoStop}%
\bibitem [{\citenamefont {Xu}\ and\ \citenamefont {Swingle}(2018)}]{otoc18}%
  \BibitemOpen
  \bibfield  {author} {\bibinfo {author} {\bibfnamefont {S.}~\bibnamefont
  {Xu}}\ and\ \bibinfo {author} {\bibfnamefont {B.}~\bibnamefont {Swingle}},\
  }\href@noop {} {\enquote {\bibinfo {title} {Accessing scrambling using matrix
  product operators},}\ } (\bibinfo {year} {2018}),\ \Eprint
  {http://arxiv.org/abs/arXiv:1802.00801} {arXiv:1802.00801} \BibitemShut
  {NoStop}%
\bibitem [{\citenamefont {Swingle}\ \emph {et~al.}(2016)\citenamefont
  {Swingle}, \citenamefont {Bentsen}, \citenamefont {Schleier-Smith},\ and\
  \citenamefont {Hayden}}]{prop1}%
  \BibitemOpen
  \bibfield  {author} {\bibinfo {author} {\bibfnamefont {B.}~\bibnamefont
  {Swingle}}, \bibinfo {author} {\bibfnamefont {G.}~\bibnamefont {Bentsen}},
  \bibinfo {author} {\bibfnamefont {M.}~\bibnamefont {Schleier-Smith}}, \ and\
  \bibinfo {author} {\bibfnamefont {P.}~\bibnamefont {Hayden}},\ }\href
  {\doibase 10.1103/PhysRevA.94.040302} {\bibfield  {journal} {\bibinfo
  {journal} {Phys. Rev. A}\ }\textbf {\bibinfo {volume} {94}},\ \bibinfo
  {pages} {040302} (\bibinfo {year} {2016})}\BibitemShut {NoStop}%
\bibitem [{\citenamefont {Zhu}\ \emph {et~al.}(2016)\citenamefont {Zhu},
  \citenamefont {Hafezi},\ and\ \citenamefont {Grover}}]{prop2}%
  \BibitemOpen
  \bibfield  {author} {\bibinfo {author} {\bibfnamefont {G.}~\bibnamefont
  {Zhu}}, \bibinfo {author} {\bibfnamefont {M.}~\bibnamefont {Hafezi}}, \ and\
  \bibinfo {author} {\bibfnamefont {T.}~\bibnamefont {Grover}},\ }\href
  {\doibase 10.1103/PhysRevA.94.062329} {\bibfield  {journal} {\bibinfo
  {journal} {Phys. Rev. A}\ }\textbf {\bibinfo {volume} {94}},\ \bibinfo
  {pages} {062329} (\bibinfo {year} {2016})}\BibitemShut {NoStop}%
\bibitem [{\citenamefont {Yao}\ \emph {et~al.}(2016)\citenamefont {Yao},
  \citenamefont {Grusdt}, \citenamefont {Swingle}, \citenamefont {Lukin},
  \citenamefont {Stamper-Kurn}, \citenamefont {Moore},\ and\ \citenamefont
  {Demler}}]{prop3}%
  \BibitemOpen
  \bibfield  {author} {\bibinfo {author} {\bibfnamefont {N.~Y.}\ \bibnamefont
  {Yao}}, \bibinfo {author} {\bibfnamefont {F.}~\bibnamefont {Grusdt}},
  \bibinfo {author} {\bibfnamefont {B.}~\bibnamefont {Swingle}}, \bibinfo
  {author} {\bibfnamefont {M.~D.}\ \bibnamefont {Lukin}}, \bibinfo {author}
  {\bibfnamefont {D.~M.}\ \bibnamefont {Stamper-Kurn}}, \bibinfo {author}
  {\bibfnamefont {J.~E.}\ \bibnamefont {Moore}}, \ and\ \bibinfo {author}
  {\bibfnamefont {E.~A.}\ \bibnamefont {Demler}},\ }\href@noop {} {\enquote
  {\bibinfo {title} {Interferometric approach to probing fast scrambling},}\ }
  (\bibinfo {year} {2016}),\ \Eprint {http://arxiv.org/abs/arXiv:1607.01801}
  {arXiv:1607.01801} \BibitemShut {NoStop}%
\bibitem [{\citenamefont {Yunger~Halpern}(2017)}]{prop4}%
  \BibitemOpen
  \bibfield  {author} {\bibinfo {author} {\bibfnamefont {N.}~\bibnamefont
  {Yunger~Halpern}},\ }\href {\doibase 10.1103/PhysRevA.95.012120} {\bibfield
  {journal} {\bibinfo  {journal} {Phys. Rev. A}\ }\textbf {\bibinfo {volume}
  {95}},\ \bibinfo {pages} {012120} (\bibinfo {year} {2017})}\BibitemShut
  {NoStop}%
\bibitem [{\citenamefont {Campisi}\ and\ \citenamefont {Goold}(2017)}]{prop5}%
  \BibitemOpen
  \bibfield  {author} {\bibinfo {author} {\bibfnamefont {M.}~\bibnamefont
  {Campisi}}\ and\ \bibinfo {author} {\bibfnamefont {J.}~\bibnamefont
  {Goold}},\ }\href {\doibase 10.1103/PhysRevE.95.062127} {\bibfield  {journal}
  {\bibinfo  {journal} {Phys. Rev. E}\ }\textbf {\bibinfo {volume} {95}},\
  \bibinfo {pages} {062127} (\bibinfo {year} {2017})}\BibitemShut {NoStop}%
\bibitem [{\citenamefont {Yunger~Halpern}\ \emph {et~al.}(2018)\citenamefont
  {Yunger~Halpern}, \citenamefont {Swingle},\ and\ \citenamefont
  {Dressel}}]{prop6}%
  \BibitemOpen
  \bibfield  {author} {\bibinfo {author} {\bibfnamefont {N.}~\bibnamefont
  {Yunger~Halpern}}, \bibinfo {author} {\bibfnamefont {B.}~\bibnamefont
  {Swingle}}, \ and\ \bibinfo {author} {\bibfnamefont {J.}~\bibnamefont
  {Dressel}},\ }\href {\doibase 10.1103/PhysRevA.97.042105} {\bibfield
  {journal} {\bibinfo  {journal} {Phys. Rev. A}\ }\textbf {\bibinfo {volume}
  {97}},\ \bibinfo {pages} {042105} (\bibinfo {year} {2018})}\BibitemShut
  {NoStop}%
\bibitem [{\citenamefont {Yoshida}\ and\ \citenamefont {Kitaev}(2017)}]{prop7}%
  \BibitemOpen
  \bibfield  {author} {\bibinfo {author} {\bibfnamefont {B.}~\bibnamefont
  {Yoshida}}\ and\ \bibinfo {author} {\bibfnamefont {A.}~\bibnamefont
  {Kitaev}},\ }\href@noop {} {\enquote {\bibinfo {title} {Efficient decoding
  for the hayden-preskill protocol},}\ } (\bibinfo {year} {2017}),\ \Eprint
  {http://arxiv.org/abs/arXiv:1710.03363} {arXiv:1710.03363} \BibitemShut
  {NoStop}%
\bibitem [{\citenamefont {Swingle}\ and\ \citenamefont
  {Yunger~Halpern}(2018)}]{prop8}%
  \BibitemOpen
  \bibfield  {author} {\bibinfo {author} {\bibfnamefont {B.}~\bibnamefont
  {Swingle}}\ and\ \bibinfo {author} {\bibfnamefont {N.}~\bibnamefont
  {Yunger~Halpern}},\ }\href {\doibase 10.1103/PhysRevA.97.062113} {\bibfield
  {journal} {\bibinfo  {journal} {Phys. Rev. A}\ }\textbf {\bibinfo {volume}
  {97}},\ \bibinfo {pages} {062113} (\bibinfo {year} {2018})}\BibitemShut
  {NoStop}%
\bibitem [{\citenamefont {Dressel}\ \emph {et~al.}(2018)\citenamefont
  {Dressel}, \citenamefont {Gonz\'alez~Alonso}, \citenamefont {Waegell},\ and\
  \citenamefont {Yunger~Halpern}}]{prop9}%
  \BibitemOpen
  \bibfield  {author} {\bibinfo {author} {\bibfnamefont {J.}~\bibnamefont
  {Dressel}}, \bibinfo {author} {\bibfnamefont {J.~R.}\ \bibnamefont
  {Gonz\'alez~Alonso}}, \bibinfo {author} {\bibfnamefont {M.}~\bibnamefont
  {Waegell}}, \ and\ \bibinfo {author} {\bibfnamefont {N.}~\bibnamefont
  {Yunger~Halpern}},\ }\href {\doibase 10.1103/PhysRevA.98.012132} {\bibfield
  {journal} {\bibinfo  {journal} {Phys. Rev. A}\ }\textbf {\bibinfo {volume}
  {98}},\ \bibinfo {pages} {012132} (\bibinfo {year} {2018})}\BibitemShut
  {NoStop}%
\bibitem [{\citenamefont {G{\"a}rttner}\ \emph {et~al.}(2017)\citenamefont
  {G{\"a}rttner}, \citenamefont {Bohnet}, \citenamefont {Safavi-Naini},
  \citenamefont {Wall}, \citenamefont {Bollinger},\ and\ \citenamefont
  {Rey}}]{exp1}%
  \BibitemOpen
  \bibfield  {author} {\bibinfo {author} {\bibfnamefont {M.}~\bibnamefont
  {G{\"a}rttner}}, \bibinfo {author} {\bibfnamefont {J.~G.}\ \bibnamefont
  {Bohnet}}, \bibinfo {author} {\bibfnamefont {A.}~\bibnamefont
  {Safavi-Naini}}, \bibinfo {author} {\bibfnamefont {M.~L.}\ \bibnamefont
  {Wall}}, \bibinfo {author} {\bibfnamefont {J.~J.}\ \bibnamefont {Bollinger}},
  \ and\ \bibinfo {author} {\bibfnamefont {A.~M.}\ \bibnamefont {Rey}},\
  }\href@noop {} {\bibfield  {journal} {\bibinfo  {journal} {Nature Physics}\
  }\textbf {\bibinfo {volume} {13}},\ \bibinfo {pages} {781} (\bibinfo {year}
  {2017})}\BibitemShut {NoStop}%
\bibitem [{\citenamefont {Li}\ \emph {et~al.}(2017)\citenamefont {Li},
  \citenamefont {Fan}, \citenamefont {Wang}, \citenamefont {Ye}, \citenamefont
  {Zeng}, \citenamefont {Zhai}, \citenamefont {Peng},\ and\ \citenamefont
  {Du}}]{exp2}%
  \BibitemOpen
  \bibfield  {author} {\bibinfo {author} {\bibfnamefont {J.}~\bibnamefont
  {Li}}, \bibinfo {author} {\bibfnamefont {R.}~\bibnamefont {Fan}}, \bibinfo
  {author} {\bibfnamefont {H.}~\bibnamefont {Wang}}, \bibinfo {author}
  {\bibfnamefont {B.}~\bibnamefont {Ye}}, \bibinfo {author} {\bibfnamefont
  {B.}~\bibnamefont {Zeng}}, \bibinfo {author} {\bibfnamefont {H.}~\bibnamefont
  {Zhai}}, \bibinfo {author} {\bibfnamefont {X.}~\bibnamefont {Peng}}, \ and\
  \bibinfo {author} {\bibfnamefont {J.}~\bibnamefont {Du}},\ }\href {\doibase
  10.1103/PhysRevX.7.031011} {\bibfield  {journal} {\bibinfo  {journal} {Phys.
  Rev. X}\ }\textbf {\bibinfo {volume} {7}},\ \bibinfo {pages} {031011}
  (\bibinfo {year} {2017})}\BibitemShut {NoStop}%
\bibitem [{\citenamefont {Wei}\ \emph {et~al.}(2018)\citenamefont {Wei},
  \citenamefont {Ramanathan},\ and\ \citenamefont {Cappellaro}}]{exp3}%
  \BibitemOpen
  \bibfield  {author} {\bibinfo {author} {\bibfnamefont {K.~X.}\ \bibnamefont
  {Wei}}, \bibinfo {author} {\bibfnamefont {C.}~\bibnamefont {Ramanathan}}, \
  and\ \bibinfo {author} {\bibfnamefont {P.}~\bibnamefont {Cappellaro}},\
  }\href {\doibase 10.1103/PhysRevLett.120.070501} {\bibfield  {journal}
  {\bibinfo  {journal} {Phys. Rev. Lett.}\ }\textbf {\bibinfo {volume} {120}},\
  \bibinfo {pages} {070501} (\bibinfo {year} {2018})}\BibitemShut {NoStop}%
\bibitem [{\citenamefont {Meier}\ \emph {et~al.}(2017)\citenamefont {Meier},
  \citenamefont {Ang'ong'a}, \citenamefont {An},\ and\ \citenamefont
  {Gadway}}]{exp4}%
  \BibitemOpen
  \bibfield  {author} {\bibinfo {author} {\bibfnamefont {E.~J.}\ \bibnamefont
  {Meier}}, \bibinfo {author} {\bibfnamefont {J.}~\bibnamefont {Ang'ong'a}},
  \bibinfo {author} {\bibfnamefont {F.~A.}\ \bibnamefont {An}}, \ and\ \bibinfo
  {author} {\bibfnamefont {B.}~\bibnamefont {Gadway}},\ }\href@noop {}
  {\enquote {\bibinfo {title} {Exploring quantum signatures of chaos on a
  floquet synthetic lattice},}\ } (\bibinfo {year} {2017}),\ \Eprint
  {http://arxiv.org/abs/arXiv:1705.06714} {arXiv:1705.06714} \BibitemShut
  {NoStop}%
\bibitem [{\citenamefont {Dicke}(1954)}]{dicke}%
  \BibitemOpen
  \bibfield  {author} {\bibinfo {author} {\bibfnamefont {R.~H.}\ \bibnamefont
  {Dicke}},\ }\href {\doibase 10.1103/PhysRev.93.99} {\bibfield  {journal}
  {\bibinfo  {journal} {Phys. Rev.}\ }\textbf {\bibinfo {volume} {93}},\
  \bibinfo {pages} {99} (\bibinfo {year} {1954})}\BibitemShut {NoStop}%
\bibitem [{\citenamefont {Kirton}\ \emph {et~al.}(2018)\citenamefont {Kirton},
  \citenamefont {Roses}, \citenamefont {Keeling},\ and\ \citenamefont
  {Torre}}]{dmrev}%
  \BibitemOpen
  \bibfield  {author} {\bibinfo {author} {\bibfnamefont {P.}~\bibnamefont
  {Kirton}}, \bibinfo {author} {\bibfnamefont {M.~M.}\ \bibnamefont {Roses}},
  \bibinfo {author} {\bibfnamefont {J.}~\bibnamefont {Keeling}}, \ and\
  \bibinfo {author} {\bibfnamefont {E.~G.~D.}\ \bibnamefont {Torre}},\
  }\href@noop {} {\enquote {\bibinfo {title} {Introduction to the dicke model:
  from equilibrium to nonequilibrium, and vice versa},}\ } (\bibinfo {year}
  {2018}),\ \Eprint {http://arxiv.org/abs/arXiv:1805.09828} {arXiv:1805.09828}
  \BibitemShut {NoStop}%
\bibitem [{\citenamefont {Emary}\ and\ \citenamefont
  {Brandes}(2003{\natexlab{a}})}]{dc1}%
  \BibitemOpen
  \bibfield  {author} {\bibinfo {author} {\bibfnamefont {C.}~\bibnamefont
  {Emary}}\ and\ \bibinfo {author} {\bibfnamefont {T.}~\bibnamefont
  {Brandes}},\ }\href {\doibase 10.1103/PhysRevLett.90.044101} {\bibfield
  {journal} {\bibinfo  {journal} {Phys. Rev. Lett.}\ }\textbf {\bibinfo
  {volume} {90}},\ \bibinfo {pages} {044101} (\bibinfo {year}
  {2003}{\natexlab{a}})}\BibitemShut {NoStop}%
\bibitem [{\citenamefont {Emary}\ and\ \citenamefont
  {Brandes}(2003{\natexlab{b}})}]{dc2}%
  \BibitemOpen
  \bibfield  {author} {\bibinfo {author} {\bibfnamefont {C.}~\bibnamefont
  {Emary}}\ and\ \bibinfo {author} {\bibfnamefont {T.}~\bibnamefont
  {Brandes}},\ }\href {\doibase 10.1103/PhysRevE.67.066203} {\bibfield
  {journal} {\bibinfo  {journal} {Phys. Rev. E}\ }\textbf {\bibinfo {volume}
  {67}},\ \bibinfo {pages} {066203} (\bibinfo {year}
  {2003}{\natexlab{b}})}\BibitemShut {NoStop}%
\bibitem [{\citenamefont {Ku\ifmmode~\acute{s}\else \'{s}\fi{}}(1985)}]{dc-1}%
  \BibitemOpen
  \bibfield  {author} {\bibinfo {author} {\bibfnamefont {M.}~\bibnamefont
  {Ku\ifmmode~\acute{s}\else \'{s}\fi{}}},\ }\href {\doibase
  10.1103/PhysRevLett.54.1343} {\bibfield  {journal} {\bibinfo  {journal}
  {Phys. Rev. Lett.}\ }\textbf {\bibinfo {volume} {54}},\ \bibinfo {pages}
  {1343} (\bibinfo {year} {1985})}\BibitemShut {NoStop}%
\bibitem [{\citenamefont {Graham}\ and\ \citenamefont
  {H\"ohnerbach}(1986)}]{dc0}%
  \BibitemOpen
  \bibfield  {author} {\bibinfo {author} {\bibfnamefont {R.}~\bibnamefont
  {Graham}}\ and\ \bibinfo {author} {\bibfnamefont {M.}~\bibnamefont
  {H\"ohnerbach}},\ }\href {\doibase 10.1103/PhysRevLett.57.1378} {\bibfield
  {journal} {\bibinfo  {journal} {Phys. Rev. Lett.}\ }\textbf {\bibinfo
  {volume} {57}},\ \bibinfo {pages} {1378} (\bibinfo {year}
  {1986})}\BibitemShut {NoStop}%
\bibitem [{\citenamefont {Altland}\ and\ \citenamefont
  {Haake}(2012{\natexlab{a}})}]{dc3}%
  \BibitemOpen
  \bibfield  {author} {\bibinfo {author} {\bibfnamefont {A.}~\bibnamefont
  {Altland}}\ and\ \bibinfo {author} {\bibfnamefont {F.}~\bibnamefont
  {Haake}},\ }\href {\doibase 10.1103/PhysRevLett.108.073601} {\bibfield
  {journal} {\bibinfo  {journal} {Phys. Rev. Lett.}\ }\textbf {\bibinfo
  {volume} {108}},\ \bibinfo {pages} {073601} (\bibinfo {year}
  {2012}{\natexlab{a}})}\BibitemShut {NoStop}%
\bibitem [{\citenamefont {Altland}\ and\ \citenamefont
  {Haake}(2012{\natexlab{b}})}]{dc4}%
  \BibitemOpen
  \bibfield  {author} {\bibinfo {author} {\bibfnamefont {A.}~\bibnamefont
  {Altland}}\ and\ \bibinfo {author} {\bibfnamefont {F.}~\bibnamefont
  {Haake}},\ }\href {http://stacks.iop.org/1367-2630/14/i=7/a=073011}
  {\bibfield  {journal} {\bibinfo  {journal} {New Journal of Physics}\ }\textbf
  {\bibinfo {volume} {14}},\ \bibinfo {pages} {073011} (\bibinfo {year}
  {2012}{\natexlab{b}})}\BibitemShut {NoStop}%
\bibitem [{\citenamefont {Bakemeier}\ \emph {et~al.}(2013)\citenamefont
  {Bakemeier}, \citenamefont {Alvermann},\ and\ \citenamefont {Fehske}}]{dc5}%
  \BibitemOpen
  \bibfield  {author} {\bibinfo {author} {\bibfnamefont {L.}~\bibnamefont
  {Bakemeier}}, \bibinfo {author} {\bibfnamefont {A.}~\bibnamefont
  {Alvermann}}, \ and\ \bibinfo {author} {\bibfnamefont {H.}~\bibnamefont
  {Fehske}},\ }\href {\doibase 10.1103/PhysRevA.88.043835} {\bibfield
  {journal} {\bibinfo  {journal} {Phys. Rev. A}\ }\textbf {\bibinfo {volume}
  {88}},\ \bibinfo {pages} {043835} (\bibinfo {year} {2013})}\BibitemShut
  {NoStop}%
\bibitem [{\citenamefont {Bastarrachea-Magnani}\ \emph
  {et~al.}(2015)\citenamefont {Bastarrachea-Magnani}, \citenamefont
  {L{\'o}pez-del Carpio}, \citenamefont {Lerma-Hern{\'a}ndez},\ and\
  \citenamefont {Hirsch}}]{dc6}%
  \BibitemOpen
  \bibfield  {author} {\bibinfo {author} {\bibfnamefont {M.~A.}\ \bibnamefont
  {Bastarrachea-Magnani}}, \bibinfo {author} {\bibfnamefont {B.}~\bibnamefont
  {L{\'o}pez-del Carpio}}, \bibinfo {author} {\bibfnamefont {S.}~\bibnamefont
  {Lerma-Hern{\'a}ndez}}, \ and\ \bibinfo {author} {\bibfnamefont {J.~G.}\
  \bibnamefont {Hirsch}},\ }\href@noop {} {\bibfield  {journal} {\bibinfo
  {journal} {Physica Scripta}\ }\textbf {\bibinfo {volume} {90}},\ \bibinfo
  {pages} {068015} (\bibinfo {year} {2015})}\BibitemShut {NoStop}%
\bibitem [{\citenamefont {Ch\'avez-Carlos}\ \emph
  {et~al.}(2016{\natexlab{a}})\citenamefont {Ch\'avez-Carlos}, \citenamefont
  {Bastarrachea-Magnani}, \citenamefont {Lerma-Hern\'andez},\ and\
  \citenamefont {Hirsch}}]{dc7}%
  \BibitemOpen
  \bibfield  {author} {\bibinfo {author} {\bibfnamefont {J.}~\bibnamefont
  {Ch\'avez-Carlos}}, \bibinfo {author} {\bibfnamefont {M.~A.}\ \bibnamefont
  {Bastarrachea-Magnani}}, \bibinfo {author} {\bibfnamefont {S.}~\bibnamefont
  {Lerma-Hern\'andez}}, \ and\ \bibinfo {author} {\bibfnamefont {J.~G.}\
  \bibnamefont {Hirsch}},\ }\href {\doibase 10.1103/PhysRevE.94.022209}
  {\bibfield  {journal} {\bibinfo  {journal} {Phys. Rev. E}\ }\textbf {\bibinfo
  {volume} {94}},\ \bibinfo {pages} {022209} (\bibinfo {year}
  {2016}{\natexlab{a}})}\BibitemShut {NoStop}%
\bibitem [{\citenamefont {Ch\'avez-Carlos}\ \emph
  {et~al.}(2016{\natexlab{b}})\citenamefont {Ch\'avez-Carlos}, \citenamefont
  {Bastarrachea-Magnani}, \citenamefont {Lerma-Hern\'andez},\ and\
  \citenamefont {Hirsch}}]{dc8}%
  \BibitemOpen
  \bibfield  {author} {\bibinfo {author} {\bibfnamefont {J.}~\bibnamefont
  {Ch\'avez-Carlos}}, \bibinfo {author} {\bibfnamefont {M.~A.}\ \bibnamefont
  {Bastarrachea-Magnani}}, \bibinfo {author} {\bibfnamefont {S.}~\bibnamefont
  {Lerma-Hern\'andez}}, \ and\ \bibinfo {author} {\bibfnamefont {J.~G.}\
  \bibnamefont {Hirsch}},\ }\href {\doibase 10.1103/PhysRevE.94.022209}
  {\bibfield  {journal} {\bibinfo  {journal} {Phys. Rev. E}\ }\textbf {\bibinfo
  {volume} {94}},\ \bibinfo {pages} {022209} (\bibinfo {year}
  {2016}{\natexlab{b}})}\BibitemShut {NoStop}%
\bibitem [{\citenamefont {Safavi-Naini}\ \emph {et~al.}(2018)\citenamefont
  {Safavi-Naini}, \citenamefont {Lewis-Swan}, \citenamefont {Bohnet},
  \citenamefont {G\"arttner}, \citenamefont {Gilmore}, \citenamefont {Jordan},
  \citenamefont {Cohn}, \citenamefont {Freericks}, \citenamefont {Rey},\ and\
  \citenamefont {Bollinger}}]{exp0}%
  \BibitemOpen
  \bibfield  {author} {\bibinfo {author} {\bibfnamefont {A.}~\bibnamefont
  {Safavi-Naini}}, \bibinfo {author} {\bibfnamefont {R.~J.}\ \bibnamefont
  {Lewis-Swan}}, \bibinfo {author} {\bibfnamefont {J.~G.}\ \bibnamefont
  {Bohnet}}, \bibinfo {author} {\bibfnamefont {M.}~\bibnamefont {G\"arttner}},
  \bibinfo {author} {\bibfnamefont {K.~A.}\ \bibnamefont {Gilmore}}, \bibinfo
  {author} {\bibfnamefont {J.~E.}\ \bibnamefont {Jordan}}, \bibinfo {author}
  {\bibfnamefont {J.}~\bibnamefont {Cohn}}, \bibinfo {author} {\bibfnamefont
  {J.~K.}\ \bibnamefont {Freericks}}, \bibinfo {author} {\bibfnamefont {A.~M.}\
  \bibnamefont {Rey}}, \ and\ \bibinfo {author} {\bibfnamefont {J.~J.}\
  \bibnamefont {Bollinger}},\ }\href {\doibase 10.1103/PhysRevLett.121.040503}
  {\bibfield  {journal} {\bibinfo  {journal} {Phys. Rev. Lett.}\ }\textbf
  {\bibinfo {volume} {121}},\ \bibinfo {pages} {040503} (\bibinfo {year}
  {2018})}\BibitemShut {NoStop}%
\bibitem [{\citenamefont {Tsvelik}(2007)}]{mj1}%
  \BibitemOpen
  \bibfield  {author} {\bibinfo {author} {\bibfnamefont {A.}~\bibnamefont
  {Tsvelik}},\ }\href {https://books.google.com/books?id=qm2\_AwAAQBAJ} {\emph
  {\bibinfo {title} {Quantum Field Theory in Condensed Matter Physics}}}\
  (\bibinfo  {publisher} {Cambridge University Press},\ \bibinfo {year}
  {2007})\BibitemShut {NoStop}%
\bibitem [{\citenamefont {Shnirman}\ and\ \citenamefont {Makhlin}(2003)}]{mj2}%
  \BibitemOpen
  \bibfield  {author} {\bibinfo {author} {\bibfnamefont {A.}~\bibnamefont
  {Shnirman}}\ and\ \bibinfo {author} {\bibfnamefont {Y.}~\bibnamefont
  {Makhlin}},\ }\href {\doibase 10.1103/PhysRevLett.91.207204} {\bibfield
  {journal} {\bibinfo  {journal} {Phys. Rev. Lett.}\ }\textbf {\bibinfo
  {volume} {91}},\ \bibinfo {pages} {207204} (\bibinfo {year}
  {2003})}\BibitemShut {NoStop}%
\bibitem [{\citenamefont {Mao}\ \emph {et~al.}(2003)\citenamefont {Mao},
  \citenamefont {Coleman}, \citenamefont {Hooley},\ and\ \citenamefont
  {Langreth}}]{mj3}%
  \BibitemOpen
  \bibfield  {author} {\bibinfo {author} {\bibfnamefont {W.}~\bibnamefont
  {Mao}}, \bibinfo {author} {\bibfnamefont {P.}~\bibnamefont {Coleman}},
  \bibinfo {author} {\bibfnamefont {C.}~\bibnamefont {Hooley}}, \ and\ \bibinfo
  {author} {\bibfnamefont {D.}~\bibnamefont {Langreth}},\ }\href {\doibase
  10.1103/PhysRevLett.91.207203} {\bibfield  {journal} {\bibinfo  {journal}
  {Phys. Rev. Lett.}\ }\textbf {\bibinfo {volume} {91}},\ \bibinfo {pages}
  {207203} (\bibinfo {year} {2003})}\BibitemShut {NoStop}%
\bibitem [{\citenamefont {Dalla~Torre}\ \emph {et~al.}(2016)\citenamefont
  {Dalla~Torre}, \citenamefont {Shchadilova}, \citenamefont {Wilner},
  \citenamefont {Lukin},\ and\ \citenamefont {Demler}}]{dmj1}%
  \BibitemOpen
  \bibfield  {author} {\bibinfo {author} {\bibfnamefont {E.~G.}\ \bibnamefont
  {Dalla~Torre}}, \bibinfo {author} {\bibfnamefont {Y.}~\bibnamefont
  {Shchadilova}}, \bibinfo {author} {\bibfnamefont {E.~Y.}\ \bibnamefont
  {Wilner}}, \bibinfo {author} {\bibfnamefont {M.~D.}\ \bibnamefont {Lukin}}, \
  and\ \bibinfo {author} {\bibfnamefont {E.}~\bibnamefont {Demler}},\ }\href
  {\doibase 10.1103/PhysRevA.94.061802} {\bibfield  {journal} {\bibinfo
  {journal} {Phys. Rev. A}\ }\textbf {\bibinfo {volume} {94}},\ \bibinfo
  {pages} {061802} (\bibinfo {year} {2016})}\BibitemShut {NoStop}%
\bibitem [{\citenamefont {Shchadilova}\ \emph {et~al.}(2018)\citenamefont
  {Shchadilova}, \citenamefont {Roses}, \citenamefont {Torre}, \citenamefont
  {Lukin},\ and\ \citenamefont {Demler}}]{dmj2}%
  \BibitemOpen
  \bibfield  {author} {\bibinfo {author} {\bibfnamefont {Y.}~\bibnamefont
  {Shchadilova}}, \bibinfo {author} {\bibfnamefont {M.~M.}\ \bibnamefont
  {Roses}}, \bibinfo {author} {\bibfnamefont {E.~G.~D.}\ \bibnamefont {Torre}},
  \bibinfo {author} {\bibfnamefont {M.~D.}\ \bibnamefont {Lukin}}, \ and\
  \bibinfo {author} {\bibfnamefont {E.}~\bibnamefont {Demler}},\ }\href@noop {}
  {\enquote {\bibinfo {title} {Fermionic formalism for driven-dissipative
  multi-level systems},}\ } (\bibinfo {year} {2018}),\ \Eprint
  {http://arxiv.org/abs/arXiv:1804.03543} {arXiv:1804.03543} \BibitemShut
  {NoStop}%
\bibitem [{\citenamefont {Stanford}(2016)}]{Stanford}%
  \BibitemOpen
  \bibfield  {author} {\bibinfo {author} {\bibfnamefont {D.}~\bibnamefont
  {Stanford}},\ }\href {\doibase 10.1007/JHEP10(2016)009} {\bibfield  {journal}
  {\bibinfo  {journal} {Journal of High Energy Physics}\ }\textbf {\bibinfo
  {volume} {2016}},\ \bibinfo {pages} {9} (\bibinfo {year} {2016})}\BibitemShut
  {NoStop}%
\bibitem [{\citenamefont {Liao}\ and\ \citenamefont
  {Galitski}(2018)}]{Galitski}%
  \BibitemOpen
  \bibfield  {author} {\bibinfo {author} {\bibfnamefont {Y.}~\bibnamefont
  {Liao}}\ and\ \bibinfo {author} {\bibfnamefont {V.}~\bibnamefont
  {Galitski}},\ }\href@noop {} {\enquote {\bibinfo {title} {Non-linear sigma
  model approach to many-body quantum chaos: regularized and unregularized
  out-of-time-ordered correlators},}\ } (\bibinfo {year} {2018}),\ \Eprint
  {http://arxiv.org/abs/arXiv:1807.09799} {arXiv:1807.09799} \BibitemShut
  {NoStop}%
\bibitem [{Note2()}]{Note2}%
  \BibitemOpen
  \bibinfo {note} {Since the Boson propagator is renormalized by the
  interaction even in the limit $N\to \infty $, we used the dressed propagator
  here. However its exact form isn't important for the purpose of current
  calculations.}\BibitemShut {Stop}%
\bibitem [{Note3()}]{Note3}%
  \BibitemOpen
  \bibinfo {note} {This does not hold for single highly excited states (as
  opposed to the thermal state) with a large fixed total spin. See
  Refs.\protect \rev@citealpnum {rey,1807a}.}\BibitemShut {Stop}%
\bibitem [{\citenamefont {Fan}\ \emph {et~al.}(2014)\citenamefont {Fan},
  \citenamefont {Yang}, \citenamefont {Zhang}, \citenamefont {Ma},
  \citenamefont {Chen},\ and\ \citenamefont {Jia}}]{ext1}%
  \BibitemOpen
  \bibfield  {author} {\bibinfo {author} {\bibfnamefont {J.}~\bibnamefont
  {Fan}}, \bibinfo {author} {\bibfnamefont {Z.}~\bibnamefont {Yang}}, \bibinfo
  {author} {\bibfnamefont {Y.}~\bibnamefont {Zhang}}, \bibinfo {author}
  {\bibfnamefont {J.}~\bibnamefont {Ma}}, \bibinfo {author} {\bibfnamefont
  {G.}~\bibnamefont {Chen}}, \ and\ \bibinfo {author} {\bibfnamefont
  {S.}~\bibnamefont {Jia}},\ }\href {\doibase 10.1103/PhysRevA.89.023812}
  {\bibfield  {journal} {\bibinfo  {journal} {Phys. Rev. A}\ }\textbf {\bibinfo
  {volume} {89}},\ \bibinfo {pages} {023812} (\bibinfo {year}
  {2014})}\BibitemShut {NoStop}%
\bibitem [{\citenamefont {Gopalakrishnan}\ \emph {et~al.}(2011)\citenamefont
  {Gopalakrishnan}, \citenamefont {Lev},\ and\ \citenamefont
  {Goldbart}}]{ext2}%
  \BibitemOpen
  \bibfield  {author} {\bibinfo {author} {\bibfnamefont {S.}~\bibnamefont
  {Gopalakrishnan}}, \bibinfo {author} {\bibfnamefont {B.~L.}\ \bibnamefont
  {Lev}}, \ and\ \bibinfo {author} {\bibfnamefont {P.~M.}\ \bibnamefont
  {Goldbart}},\ }\href {\doibase 10.1103/PhysRevLett.107.277201} {\bibfield
  {journal} {\bibinfo  {journal} {Phys. Rev. Lett.}\ }\textbf {\bibinfo
  {volume} {107}},\ \bibinfo {pages} {277201} (\bibinfo {year}
  {2011})}\BibitemShut {NoStop}%
\bibitem [{\citenamefont {Strack}\ and\ \citenamefont {Sachdev}(2011)}]{ext3}%
  \BibitemOpen
  \bibfield  {author} {\bibinfo {author} {\bibfnamefont {P.}~\bibnamefont
  {Strack}}\ and\ \bibinfo {author} {\bibfnamefont {S.}~\bibnamefont
  {Sachdev}},\ }\href {\doibase 10.1103/PhysRevLett.107.277202} {\bibfield
  {journal} {\bibinfo  {journal} {Phys. Rev. Lett.}\ }\textbf {\bibinfo
  {volume} {107}},\ \bibinfo {pages} {277202} (\bibinfo {year}
  {2011})}\BibitemShut {NoStop}%
\bibitem [{\citenamefont {Gopalakrishnan}\ \emph {et~al.}(2009)\citenamefont
  {Gopalakrishnan}, \citenamefont {Lev},\ and\ \citenamefont
  {Goldbart}}]{ext4}%
  \BibitemOpen
  \bibfield  {author} {\bibinfo {author} {\bibfnamefont {S.}~\bibnamefont
  {Gopalakrishnan}}, \bibinfo {author} {\bibfnamefont {B.~L.}\ \bibnamefont
  {Lev}}, \ and\ \bibinfo {author} {\bibfnamefont {P.~M.}\ \bibnamefont
  {Goldbart}},\ }\href@noop {} {\bibfield  {journal} {\bibinfo  {journal}
  {Nature Physics}\ }\textbf {\bibinfo {volume} {5}},\ \bibinfo {pages} {845}
  (\bibinfo {year} {2009})}\BibitemShut {NoStop}%
\bibitem [{\citenamefont {Ch\'avez-Carlos}\ \emph {et~al.}(2018)\citenamefont
  {Ch\'avez-Carlos}, \citenamefont {del Carpio}, \citenamefont
  {Bastarrachea-Magnani}, \citenamefont {Stransky}, \citenamefont
  {Lerma-Hern\'andez}, \citenamefont {Santos},\ and\ \citenamefont
  {Hirsch}}]{1807a}%
  \BibitemOpen
  \bibfield  {author} {\bibinfo {author} {\bibfnamefont {J.}~\bibnamefont
  {Ch\'avez-Carlos}}, \bibinfo {author} {\bibfnamefont {B.~L.}\ \bibnamefont
  {del Carpio}}, \bibinfo {author} {\bibfnamefont {M.~A.}\ \bibnamefont
  {Bastarrachea-Magnani}}, \bibinfo {author} {\bibfnamefont {P.}~\bibnamefont
  {Stransky}}, \bibinfo {author} {\bibfnamefont {S.}~\bibnamefont
  {Lerma-Hern\'andez}}, \bibinfo {author} {\bibfnamefont {L.~F.}\ \bibnamefont
  {Santos}}, \ and\ \bibinfo {author} {\bibfnamefont {J.~G.}\ \bibnamefont
  {Hirsch}},\ }\href@noop {} {\enquote {\bibinfo {title} {Quantum and classical
  lyapunov exponents in atom-field interaction systems},}\ } (\bibinfo {year}
  {2018}),\ \Eprint {http://arxiv.org/abs/arXiv:1807.10292} {arXiv:1807.10292}
  \BibitemShut {NoStop}%
\bibitem [{\citenamefont {Lewis-Swan}\ \emph {et~al.}(2018)\citenamefont
  {Lewis-Swan}, \citenamefont {Safavi-Naini}, \citenamefont {Bollinger},\ and\
  \citenamefont {Rey}}]{rey}%
  \BibitemOpen
  \bibfield  {author} {\bibinfo {author} {\bibfnamefont {R.~J.}\ \bibnamefont
  {Lewis-Swan}}, \bibinfo {author} {\bibfnamefont {A.}~\bibnamefont
  {Safavi-Naini}}, \bibinfo {author} {\bibfnamefont {J.~J.}\ \bibnamefont
  {Bollinger}}, \ and\ \bibinfo {author} {\bibfnamefont {A.~M.}\ \bibnamefont
  {Rey}},\ }\href@noop {} {\enquote {\bibinfo {title} {Unifying fast
  scrambling, thermalization and entanglement through the measurement of fotocs
  in the dicke model},}\ } (\bibinfo {year} {2018}),\ \Eprint
  {http://arxiv.org/abs/arXiv:1808.07134} {arXiv:1808.07134} \BibitemShut
  {NoStop}%
\bibitem [{\citenamefont {Rozenbaum}\ \emph {et~al.}(2017)\citenamefont
  {Rozenbaum}, \citenamefont {Ganeshan},\ and\ \citenamefont
  {Galitski}}]{Efim1}%
  \BibitemOpen
  \bibfield  {author} {\bibinfo {author} {\bibfnamefont {E.~B.}\ \bibnamefont
  {Rozenbaum}}, \bibinfo {author} {\bibfnamefont {S.}~\bibnamefont {Ganeshan}},
  \ and\ \bibinfo {author} {\bibfnamefont {V.}~\bibnamefont {Galitski}},\
  }\href {\doibase 10.1103/PhysRevLett.118.086801} {\bibfield  {journal}
  {\bibinfo  {journal} {Phys. Rev. Lett.}\ }\textbf {\bibinfo {volume} {118}},\
  \bibinfo {pages} {086801} (\bibinfo {year} {2017})}\BibitemShut {NoStop}%
\end{thebibliography}%
\appendix
\begin{widetext}
  \section{Explicit forms of $f_{\tilde\sigma_z}(\omega,p)$ and $f_{\tilde\phi}(\omega,p)$}\label{a1}
  In this appendix we present the explicit expressions of Eqs. \eqref{m1} and \eqref{m2}. By plugging  Eq. \eqref{rmj} into Eq.\eqref{m1}, one arrives at an explicit integral equation for $f_{\tilde \sigma_z}(\omega,p)$ with double integrals over $q$ and $\Omega$. Due to delta functions in the expression of Wightman functions, both integrals can be carried out easily to get,
  \begin{align}\label{me1}
    &f_{\tilde\sigma_z}(\omega,p)=\frac{-g^4}{4N\cosh^2(\beta \tilde\omega_z/2)}G^R_{\tilde\eta}(p)G^R_{\tilde\eta}(\omega-p) \Bigg[G_{\tilde{\phi}}^R(p-\tilde\omega_z)G_{\tilde{\phi}}^R(\omega+\tilde\omega_z-p)f_{\tilde\sigma_z}(\omega,p-2\tilde\omega_z)\nonumber \\ &+G_{\tilde{\phi}}^R(p+\tilde\omega_z)G_{\tilde{\phi}}^R(\omega-\tilde\omega_z-p)f_{\tilde\sigma_z}(\omega,p+2\tilde\omega_z)
    +\Big(G_{\tilde{\phi}}^R(p-\tilde\omega_z)G_{\tilde{\phi}}^R(\omega+\tilde\omega_z-p)\nonumber \\
    &+G_{\tilde{\phi}}^R(p+\tilde\omega_z)G_{\tilde{\phi}}^R(\omega-\tilde\omega_z-p)\Big)f_{\tilde\sigma_z}(\omega,p)\Bigg]
  \end{align}

  Similarly, to obtain the explicit form of Eq.\eqref{m2}, we use the diagrammatic expression of $R_{\tilde \phi}(\omega,p,q)$ in Fig.\ref{Boson_rung.pdf} to find its algebraic form in terms of Fermionic Green's functions. By plugging in this expression into Eq.\eqref{m2} and performing the integrals using the delta functions coming from Wightman functions, we get,
  
  \begin{figure}[t]
  \includegraphics[width=0.5\textwidth]{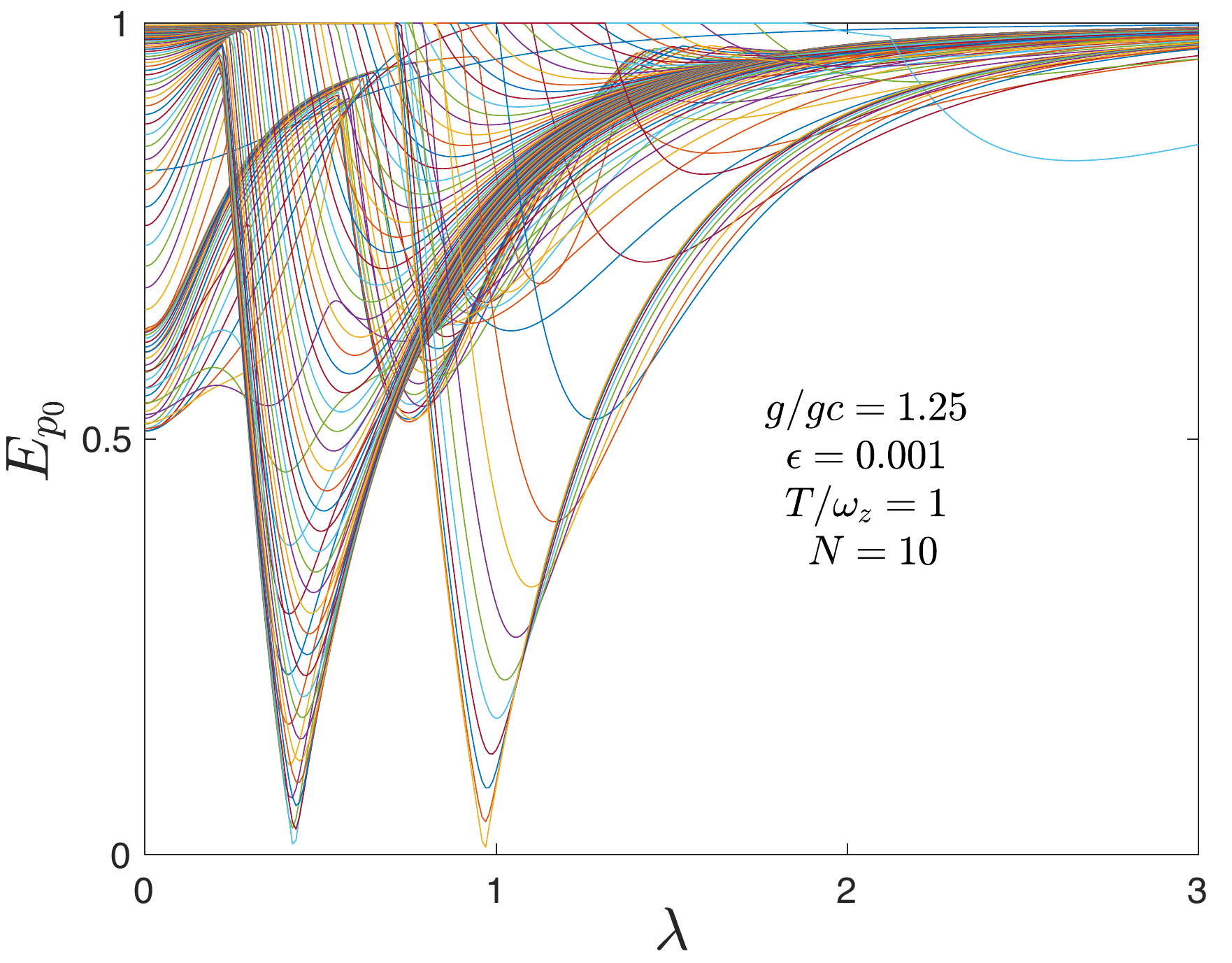}
  \caption{Smallest magnitude eigenvalues of different blocks of the matrix $M(i\lambda)$ versus $\lambda$.}
  \label{gisu}
\end{figure}

  \begin{align}\label{me2}
    &f_{\tilde\phi}(\omega,p)=\frac{-g^4}{N}G^R_{\tilde\phi} (p) G^R_{\tilde\phi} (\omega-p)\Bigg[\frac{1}{4\cosh^2(\beta\tilde\omega_z/2)}
    \Big(\cos^2(\theta)G^R_{\tilde\eta}(p-\tilde\omega_z)+4\sin^2(\theta)G^R_{\tilde{f}^\dagger}(p-\tilde\omega_z)\Big)\Big(\cos^2(\theta)\big(G^R_{\tilde{\eta}}(\omega-p+\tilde\omega_z)\nonumber\\
    &-G^R_{\tilde{\eta}}(\omega-p-\tilde\omega_z)\big)+4\sin^2(\theta)\big(G^R_{\tilde{f}}(\omega-p+\tilde\omega_z)-G^R_{\tilde{f}^\dagger}(\omega-p-\tilde\omega_z)\big)\Big)f_{\tilde\phi}(\omega,p)
    +\frac{2\sin^2(\theta)\cos^2(\theta)}{\cosh(\beta\tilde\omega_z/2)}\Big(
    \big(G^R_{\tilde{f}^\dagger}(\omega-p-\tilde\omega_z)\nonumber \\
    &-G^R_{\tilde{f}}(\omega-p)\big)G^R_{\tilde{f}}(p+\tilde\omega_z)+\big(G^R_{\tilde{f}}(\omega-p) -G^R_{\tilde{f}^\dagger}(\omega-p-\tilde\omega_z) \big)G^R_{\tilde{f}^\dagger}(p)\Big)f_{\tilde\phi}(\omega,p+\tilde\omega_z)
    \Bigg]+\big( \tilde{\omega}_z \to -\tilde{\omega}_z \big)
  \end{align}
\section{Details of computing $\lambda^{\tilde\phi}_L$}\label{a2}
Here we explain how to compute $\lambda_L^{\tilde\phi}$ in more detail. $\lambda_L^{\tilde\sigma_z}$ can be obtained in a similar way. We start by writing Eq.\eqref{me1} in the matrix form,
\begin{align}\label{mxe}
\sum_q M_{p,q}(\omega) f_q(\omega)=0,
\end{align}
where the matrix elements of $M(\omega)$ is given by,
\begin{align}\label{Mmatrix}
  M_{p,q}(\omega)=&\Bigg[1+\frac{g^4}{4N\cosh^2(\beta \tilde\omega_z/2)}G^R_{\tilde\eta}(p)G^R_{\tilde\eta}(\omega-p)\Big(G_{\tilde{\phi}}^R(p-\tilde\omega_z)G_{\tilde{\phi}}^R(\omega+\tilde\omega_z-p)+G_{\tilde{\phi}}^R(p+\tilde\omega_z)G_{\tilde{\phi}}^R(\omega-\tilde\omega_z-p)\Big)\Bigg]\delta_{p,q} \nonumber \\
  &+\frac{g^4}{4N\cosh^2(\beta \tilde\omega_z/2)}G^R_{\tilde\eta}(p)G^R_{\tilde\eta}(\omega-p)G_{\tilde{\phi}}^R(p-\tilde\omega_z)G_{\tilde{\phi}}^R(\omega+\tilde\omega_z-p)\delta_{p-2\tilde\omega_z,q}\nonumber \\
  &+\frac{g^4}{4N\cosh^2(\beta \tilde\omega_z/2)}G^R_{\tilde\eta}(p)G^R_{\tilde\eta}(\omega-p)G_{\tilde{\phi}}^R(p+\tilde\omega_z)G_{\tilde{\phi}}^R(\omega-\tilde\omega_z-p)\delta_{p+2\tilde\omega_z,q}.
\end{align}
We call $p$ and $q$ frequency indices. As was mentioned in section \ref{reslt}, we want to find largest $\lambda>0$ such that $M(i\lambda)$ has a zero eigenvalue. To this end, we probe the positive imaginary axis and compute the smallest magnitude eigenvalue of $M(i\lambda)$ in each point to find the value of $\lambda$ where this eigenvalue becomes zero. Note that due to simple form of expression \eqref{Mmatrix}, the $M$ matrix couples frequency $p$ only to itself and $p\pm 2 \tilde\omega_z$. As a consequence, $M$ can be written in a block diagonal form where each block consists of frequencies,
$$ p_n=p_0+2n\tilde\omega_z, \qquad n\in\mathbb{Z}. $$
We use $p_0$ to label each block. The block diagonal form of $M$ makes finding its eigenvalues significantly easier since we can diagonalize each block separately. A typical plot showing the smallest magnitude eigenvalue of each block($E_{p_0}$) versus $\lambda$ is given in Fig.\ref{gisu} where different lines corresponds to different $p_0$'s.

 In the main text we reported $\lambda_L$ for $\varepsilon \rightarrow 0$ ($\varepsilon$ is the imaginary part of the retarded Green's function denominator) and discarded solutions that are strongly sensitive to $\varepsilon$. However, as stated in the main text, the leading order correction to the imaginary part of the Green's functions is of the order $1/N$. In anticipation of this, we once choose a fixed $\varepsilon=\mathcal{O}(1/N)$ ($\varepsilon$ is the imaginary part of Green's function denominator) and confirm that the resulting behavior is qualitatively the same as what is reported in the main text (keeping \textit{all} solutions).

\end{widetext}

\end{document}